# Delivery of icy planetesimals to inner planets in the Proxima Centauri planetary system

Sergei I. IPATOV *

V.I. Vernadsky Institute of Geochemistry and Analytical Chemistry of RAS, Moscow, Russia
*Corresponding author. E-mail: siipatov@hotmail.com

**Abstract** -- The estimates of the delivery of icy planetesimals from the feeding zone of Proxima Centauri $c$ (with mass equal to $7m_E$, $m_E$ is the mass of the Earth) to inner planets $b$ and $d$ were made. They included the studies of the total mass of planetesimals in the feeding zone of planet $c$ and the probabilities of collisions of such planetesimals with inner planets. This total mass could be about 10-15$m_E$. It was estimated based on studies of the ratio of the mass of planetesimals ejected into hyperbolic orbits to the mass of planetesimals collided with forming planet $c$. At integration of the motion of planetesimals, the gravitational influence of planets $c$ and $b$ and the star was taken into account. In most series of calculations, planetesimals collided with planets were excluded from integrations. Based on estimates of the mass of planetesimals ejected into hyperbolic orbits, it was concluded that during the growth of the mass of planet $c$ the semi-major axis of its orbit could decrease by at least a factor of 1.5. Depending on possible gravitational scattering due to mutual encounters of planetesimals, the total mass of material delivered by planetesimals from the feeding zone of planet $c$ to planet $b$ was estimated to be between $0.002m_E$ and $0.015m_E$. Probably, the amount of water delivered to Proxima Centauri $b$ exceeded the mass of water in Earth's oceans. The amount of material delivered to planet $d$ could be a little less than that delivered to planet $b$.

# INTRODUCTION

**Proxima Centauri system**

More than 5000 exoplanets have been discovered. Some exoplanets are located in the zones where life on exoplanets could exist. The definition of habitable zones for extrasolar planetary systems is traditionally based on the conditions promoting the presence of standing bodies of liquid surface water (determined to be a conventional habitable zone), but other more refined boundaries may be considered (Kopparapu et al., 2013; Yamashiki et al., 2019). High ionizing radiation fluxes including X-ray and extreme ultraviolet emission, coronal mass ejections, and associated stellar energetic particles events can affect exoplanetary habitability conditions (Airapetian et al., 2017; Yamashiki et al., 2019). Marov and Shevchenko (2020) discussed criteria for habitability of an exoplanet (size, mass, radial distance from a host star, chemical composition of atmosphere, stability of climate). Chambers (2020) noted that liquid water is not only prerequisite for life to exist. He supposed that a number of other conditions (e.g. the surface temperature of a planet and the partial pressure of $CO_2$ in the atmosphere) exist before life can form. Underwood et al. (2003) concluded that the major effect is that a star becomes more luminous, causing a habitable zone to move outwards. Schulze-Makuch et al. (2020) supposed that extraterrestrial life could exhibit a much large variety of forms than life on the Earth, and it is impossible to measure habitability with our current knowledge.

Our nearest stellar neighbour, Proxima Centauri, is a low-mass star with spectral type dM5.5. Its mass is about 0.1221 of the solar mass. The Proxima Centauri system is an example of exoplanetary systems with exoplanets in a habitable zone. Proxima Centauri (α Centauri C) is a member of a triple star system (Lalitha et al., 2020). This system includes also a binary star, which consists of two stars with masses about the Sun's mass: α Centauri A (officially Rigil Kentaurus) and α Centauri B (officially Toliman). Currently, the distance between Proxima Centauri and Alpha Centauri AB is about 12,950 AU (astronomical units). Proxima Centauri has three planets. Masses (in the Earth's masses $m_E$) and semi-major axes (in AU) and eccentricities of orbits of planets *b*, *c,* and *d* are presented in Table 1 and were used in our calculations.

Table 1. Masses (in the Earth's masses) and semi-major axes (in AU) and eccentricities of orbits of planets *b*, *c,* and *d* in the Proxima Centauri system.

| planet | mass, $m_E$ | semi-major axis, AU | eccentricity |
|--------|-------------|---------------------|--------------|
| *b*    | 1.17        | 0.04875             | 0.11         |
| *c*    | 7           | 1.489               | 0.04         |
| *d*    | 0.29        | 0.02895             | 0            |

The figures from (Kopparapu et al., 2013, 2014) show that a habitable zone around a star with a mass equal to 0.1 solar mass is located at a distance of about 0.03 – 0.06 AU from the star. Anglada-Escudé et al. (2016) concluded that Proxima Centauri *b* with a semi-major axis of 0.05 AU lies in the centre of the classical habitable zone for Proxima. They wrote that habitability of planets like Proxima *b* - in the sense of sustaining an atmosphere and liquid water on its surface - is a matter of intense debate. The most common arguments against habitability are tidal locking, strong stellar magnetic field, strong flares, and high UV & X-ray fluxes; but none of these arguments has been proven definitive. Airapetian et al. (2017) considered the mass loss of oxygen driven by XUV fluxes from Proxima Centauri and concluded that the escape time of a 1 bar oxygen-rich atmosphere on Proxima Centauri *b* is expected to be ~10 Myr. Habitability of Proxima Centauri *b* was also discussed by Barnes et al. (2018), Del Genio et al. (2019), Howard (2018), Meadows et al. (2018), and Ribas et al. (2016).

Meng et al. (2019) studied the dynamical evolution of the Proxima Centauri system with the full equations of motion and semi-analytical models including relativistic and tidal effects. They found that the orbit of Proxima Centauri *b* is stable for the semi-major axis ranging from 0.02 to 0.1 AU and the eccentricity being less than 0.4.

The motion of exocomets in the Proxima Centauri system was studied by Schwarz et al. (2018). They assumed an Oort cloud-like distribution of inclinations $i$ of exocomets with $0 \leq i \leq 180°$. Perihelion distances of exocomets were less than 0.0485 AU, and initial eccentricities were between 0.95 and 0.9999. Initial semi-major axes $a_{exc}$ of exocomets varied in different series of calculations from 1 to 1000 AU. Schwarz et al. (2018) considered Proxima Centauri $b$ with a semi-major axis $a_b$=0.0485 AU located in a habitable zone, and Proxima Centauri $c$ with a semi-major axis $a_c$ from 0.06 AU to up to 0.3 AU (for test calculations up to 0.7 AU). The limits of the habitable zone were calculated by Schwarz et al. (2018) to be 0.045–0.119 AU. In some series of calculations, the influence of the star binary system (Alpha Centauri AB) was taken into account. The number of impacts of 5000 exocomets with Proxima Centauri $b$ was less than 10. Schwarz et al. (2018) concluded that the star binary perturbs exocomets, but does not increase the influx of exocomets (and water transport) to exoplanets. They obtained that during 2 Myr more that 3/5 of exocomets with $990 \leq a_{exc} \leq 1000$ AU reached 20 000 AU from Proxima Centauri or collided with it in calculations that took into account the gravitational influence of the star binary system.

**Models of accretion of planets**

It is considered (e.g., Carerra et al., 2015; Jansson et al., 2017; Johansen et al., 2011, 2012; Morbidelli et al., 2009) that a lot of planetesimals could form in a protoplanetary disk (e.g., as a result of a streaming instability) and some planetesimals could exceed the sizes of largest asteroids. Many authors (e.g. Chambers, 2001, 2013; Chambers and Wetherill, 1998; Cox and Lewis, 1980; Fischer and Ciesla, 2014; Haghighipour and Winter, 2014; Ipatov, 1987a,b, 1988, 1991, 1993, 2019; Marov and Ipatov, 2023; Morbidelli et al., 2012; Lykawka and Ito, 2017; Raymond et al., 2005; Safronov, 1972; Turrini, 2014; Wetherill, 1980) studied the role of planetesimals in the planet accretion. These studies were mainly based on the evolutionary models for disks of gravitating bodies, which coagulated under collisions.

In recent years, the role of pebble accretion in the formation of planets (especially of giant planets) was studied (e.g., Lambrechts and Johansen, 2012, 2014; Morbidelli, 2020; Morbidelli et al., 2015). Pebbles accrete onto big bodies in gas-rich enviroments. Lambrechts and Johansen (2014) concluded that the ratio of the number of pebbles accreted by a core at 5 AU over the number of those drifted by was about 20 per cent, and after the formation of the giant planets, a sufficient amount of solids was left for a planetesimal disk. Valletta and Helled (2020) noted that pebble accretion is faster compared to the planetesimal case. The pebble accretion stops when a planet reaches so called pebble isolation mass.

In our Solar System, asteroids and trans-Neptunian objects are examples of former planetesimals. For the Solar System, the delivery of planetesimals from the zone of the giant planets to the Earth was studied e.g. by Morbidelli et al. (2000), Levison et al. (2001), Ipatov and Mather (2003, 2004a-b), Marov and Ipatov (2018, 2023), Ipatov (2020). Morbidelli et al. (2000) considered test particles with initial semi-major axes of their orbits ranging from 4.5 to 9 AU and eccentricities and inclinations equaled to zero. Levison et al. (2001) studied the delivery of particles with initial semi-major axes of particles' orbits from 15 to 32 AU. The delivery of bodies from the outer part of the main asteroid belt to the Earth was studied in (Morbidelli et al., 2000, 2012; Petit et al., 2001; Lunine et al., 2003; Raymond et al., 2004; O'Brien et al., 2014; Ipatov, 2021b).

**Problems considered in the paper**

The gas-free stage of formation of the Proxima Centauri planetary system is considered in this paper. The main aim of the studies was to estimate the delivery of icy planetesimals from the feeding zone of planet Proxima Centauri $c$ to planets Proxima Centauri $b$ and $d$.

Initial data and algorithms used for calculations are discussed in the section "Initial data and algorithms used for calculations". Initial eccentricities of planetesimals equaled to 0.02 or 0.15 and were much smaller that the eccentricities of exocomets (>0.95) considered by Schwarz et al. (2018). I considered the evolution of orbits of planetesimals during formation of planet *c*, but not only Oort type orbits as the above authors. Below the model of scattering of planetesimals (or smaller bodies) initially located in the feeding zone of the outer exoplanet Proxima Centauri *c* during its growth is studied. According to (Kervella et al., 2020; Benedict and McArthur, 2020), the value of the semi-major axis of orbit of planet *c* was considered to be equal to $a_c$=1.489 AU and differs much from the values used by Schwarz et al. (2018). The Proxima Centauri planetary system is an example of a planetary system with one dominant planet.

Most of calculations have been made for the present masses of the planets. However, some calculations were also made for the mass of planet *c* equal to 0.5 or 0.1 of its present mass. For the considered model, it was supposed that some planetesimals still moved near the orbit of Proxima Centauri *c* or could cross the orbit after some time. Some planetesimals or smaller bodies, that had not been located at some moment of time in the vicinity of the orbit of exoplanet *c* (e.g., had been formed at larger or smaller distances from the host star), could get in the vicinity with time. One of the factors that allowed some planetesimals to change their distances from the star was gravitational scattering due to their mutual encounters. After planetesimals-bodies reached the vicinity of the orbit of planet *c*, their further motion can be studied by the calculation model presented in this paper. Exoplanet *c* is located beyond the water-ice condensation line, and planet *b* could be located in the habitable zone. It was assumed that planet *b* formed dry, while planetesimals near planet *c* were volatile-rich, independent of how they formed. Planetesimals from the outer region could deliver water and volatiles to the inner planets *b* and *d*. For the initial data for the considered model, there is no much difference how and where Proxima Centauri planets acquired a considerable fraction of their masses, and what fractions of their masses have been delivered by planetesimals or by pebbles. The formation of growing planets and planetesimals before these initial data had been reached are not discussed.

The studies presented below in the section "Results of calculations" included the calculations of the probabilities of collisions of planetesimals with Proxima Centauri *c*, the probabilities of their ejection into hyperbolic orbits, and the probabilities of collisions of planetesimals with Proxima Centauri *b* and *d*. Special attention is paid to the dependence of the obtained results on initial values of semi-major axes of orbits of planetesimals and their initial eccentricities. In the main series *C* of calculations, planetesimals were removed from integrations if they collided with planets or the star or were ejected into hyperbolic orbits.

In contrast to the series *C*, in abstracts (Ipatov, 2021a) the studies of evolution of orbits of the planetesimals which initial orbits were close to the orbit of planet *c* were presented shortly for other series of calculations. In those calculations, planets were considered as point masses, and planetesimals were not excluded from the integration when they passed within the physical radius of a planet. In those series of calculations, denoted below as the *MP* and *F* series, the probabilities of collisions of planetesimals with planets were calculated based on the arrays of orbital elements of planetesimals during considered time intervals. For these series, the models *A* and *B* that are discussed in Appendix A show the interval of the values of the probabilities. Both models use the same sum of probabilities calculated based on the arrays of orbital elements of all planetesimals, but they estimate in different ways how to correct these probabilities in order to take into account that the real number of planetesimals at a current time must differ from that obtained at the *MP* or *F* integrations because the number of planetesimals in a disk must also decrease due to collisions of planetesimals with planets. For the *MP* series, initial masses and orbits of planets were the same as those for the series *C*. For the *F* series, the mass of planet *c* was equal to 12$m_E$, as it was considered by astronomers a few years ago. The values of the probabilities of collisions of planetesimals with planets for the *MP* and *F* series are discussed in Appendix B.

The calculations for the *MP* and *F* series could overestimate the gravitational influence of planets because some retaining planetesimals could pass within the physical radius of a planet. However, it is interesting to find the difference between the probabilities of collisions of planetesimals with planets for the series *C* and *MP*. It was obtained that such probabilities differed by a factor not more than 2. As quite different models of calculations, and also the calculations with different time steps of integration, allows one to make similar conclusions about the delivery of icy bodies to inner planets, it is possible to suppose that the conclusions could be true. Calculations for the *MP* and *F* series allowed us to make estimates of the delivery of icy planetesimals to planet *d*, which was not included in integrations for the series *C*. The results for the series *F* allow one to estimate how a greater mass of planet *c* can increase the delivery of material to inner planets. The approach similar to the *MP* and *F* calculations can be useful for the case of very small probabilities of collisions of planetesimals with some planet. In such case, the model that includes collisions with that planet in the integration (similar to the *C* calculations) needs to consider a huge number of planetesimals. I started calculations from the *MP* series because at first I did not know how small could be the probabilities of collisions of considered planetesimals with planet *b*.

The considered models correspond to the late stage of the growth of exoplanet *c*. They can be used for estimates of scattering of planetesimals and accumulation of planets in other planetary systems, especially in those in which the mass of one planet is greater (by at least a factor of several) than masses of other planets. The calculations have been made for the model that takes into account only the gravitational influence of the star and the planets. They show the role of planets in the evolution of disks of planetesimals. The evolution of disks of planetesimals depends much on distributions of sizes of planetesimals, which could be different for different models. The considered models can be useful for estimates of the dependence of the evolution of disks of planetesimals on initial data, and for understanding which initial data would be better to choose for simulations of the evolution of the disk consisted of a much greater number of initial planetesimals.

Based on the probabilities of collisions of planetesimals with planets and on other results of calculations presented in the section "Results of calculations", several conclusions have been made in the last section "Variations of the orbit of growing planet *c* and the delivery of icy planetesimals to planets *b* and *d*". At first, the total mass of planetesimals in the feeding zone of Proxima Centauri *c* was estimated. This total mass is close to the sum of masses of planetesimals encorporated into planet *c* and the sum of masses of planetesimals ejected into hyperbolic orbits. Only a small fraction of material was delivered to the star and inner planets. Estimates of the total mass of ejected planetesimals were based on the calculations of the ratio of the mass of planetesimals ejected into hyperbolic orbits to the mass of planetesimals collided with forming planet *c*. The ratio was estimated for a few values of a mass of growing planet *c*. Then the estimates of the decrease in a semi-major axis of an orbit of growing planet *c* were made based on the integral of energy, the variations in the mass of the planet, and the total mass of ejected planetesimals. In two last subsections, the delivery of icy planetesimals from the feeding zone of planet *c* to inner planets *b* and *d* is considered. This delivery is compared with the delivery of icy planetesimals from the feeding zone of the giant planets to the Earth. The total mass of icy planetesimals delivered from the feeding zone of planet *c* to planets *b* and *d* depended on the total mass of planetesimals in the feeding zone of planet *c* and on the probabilities of collisions of such planetesimals with planets *b* and *d*.

## INITIAL DATA AND ALGORITHMS USED FOR CALCULATIONS

**The main series *C* of calculations with removal of planetesimals that collided with planets**

The model of motion of planetesimals in the Proxima Centauri planetary system was studied. Besides the star with a mass equal to 0.122 of the solar mass, exoplanets *b* and *c* were considered in the calculations. Initial data for exoplanets were taken from https://en.wikipedia.org/wiki/Proxima_Centauri. For the main **series *C*** of calculations, the data were taken from this website in June 2020 and correspond to observations by Benedict and McArthur (2020). The semi-major axes and eccentricities of orbits and masses ($m_b$ and $m_c$) of Proxima Centauri *b* and *c* are presented in Table 1. The inclinations $i_b$ and $i_c$ of the orbits of the exoplanets were supposed to be equal to 0. The calculations were made also for growing planet Proxima Centauri *c* with a mass equal to $m_{c05}=0.5m_c=3.5m_E$ or $m_{c01}=0.1m_c=0.7m_E$ (and with the same orbit as that for planet *c*). The main aim of the studies with different values of the mass of planet *c* was to understand how the obtained results depend on this mass in order to make some estimates for growing planet *c* (e.g. of the evolution of its orbit and the total mass of planetesimals in its feeding zone).

Planetesimals were initially located in some vicinity of the orbit of Proxima Centauri *c*. In each calculation variant, initial semi-major axes $a_o$ of orbits of planetesimals were in the range from $a_{min}$ to $a_{max}=a_{min}+0.1$ AU. The considered values of $a_{min}$ varied from 0.9 to 2.2 AU. The considered time interval *T* often was not less than 100 Myr. For variants with $a_{min}$ from 1.2 to 1.7 AU and $m_c=7m_E$, calculations were typically made for a few hundreds of million years (up to 1 Gyr). The number of planetesimals with $a_o$ was proportional to $a_o^{1/2}$, i.e. surface density was proportional to $a_o^{-1/2}$. For the (*i*+1)th planetesimal, the value of $a_o$ was calculated with the use of the formula $a_{o(i+1)}=(a_{oi}^2+[(a_{min}+d_a)^2-a_{min}^2]/N_o)^{1/2}$, where $a_{oi}$ is the value of $a_o$ for *i*th planetesimal, $d_a=0.1$ AU, $N_o=250$ is the number of planetesimals in one variant. Initial eccentricities $e_o$ of orbits of planetesimals were equal to 0.02 or 0.15 for the main series *C* of calculations. Initial inclinations of orbits of the planetesimals were equal to $e_o/2$ rad. Small eccentricities of orbits of planetesimals were at the stage of their formation. Greater eccentricities could be a result of the gravitational scattering due to mutual encounters of planetesimals. Calculations showed (Ipatov, 1987b, 1993) that even for the feeding zone of the terrestrial planets the mean eccentricity of planetesimals could exceed 0.2 during evolution and reach 0.4 at the late stages. I studied the scattering of planetesimals at the late stages of formation of exoplanets. Therefore, calculations have been made not only for small eccentricities of planetesimals, but also at $e_o=0.15$. Exclusive for a few cases that are noted specially, all the results presented below (e.g. in all figures and in Tables 2-4) are for the model *C*.

The motion of planetesimals and exoplanets was calculated with the use of the RMVS3 version of the symplectic code from (Levison and Duncan, 1994). Many scientists used this code until now. Frantseva et al (2022), who used the code, noted that a time step of integration is reducing considerably at a distance less than 3.5 Hill radii. Ipatov and Mather (2004a,b) showed that the symplectic code and the BULSTO code (Bulirsh and Stoer, 1966) gave similar results for studies of the orbital evolution of bodies in the Solar System. In most calculations, the integration time step $t_s$ equaled to 1 day. At $t_s=1$ day, the considered time interval was not less than 100 Myr (if the evolution of a disk of planetesimals had not finished earlier). The integrations with $t_s$ equaled to 0.1, 0.2, 0.5, and 2 days also have been made. The obtained results (presented e.g. in Figs. 1-4) were about the same for different considered $t_s$.

The gravitational influence of the star and planets *b* and *c* was taken into account. The gravitational scattering due to mutual encounters of planetesimals and their mutual collisions were not taken into account at the integration. The planetesimals that collided with the star (or planets in the series *C*) or reached 1200 AU from the star were excluded from integration. The motion of most (83-92 per cent) planetesimals from 500 to 1200 AU took less than 1 Myr, so in some variants calculations stoped when planetesimals reached 500 AU from the star. According to Schwarz et al. (2018), the radius of the Hill sphere of Proxima Centauri equals to 1200 AU. Probably, the planetesimals that left this sphere have very small chances to return to this sphere later.

Schwarz et al. (2018) calculated the motion of highly eccentrical orbits of exocomets including those near the edge of the Hill sphere. They showed that the probabilities of collisions of such exocomets with planets for the model that considers only the gravitational influence of the star and the planets are about the same as those for the model that took into account also the gravitational influence of the star binary system (Alpha Centauri AB). The difference in the probabilities should be much smaller for the orbits that were located much closer to the star. Therefore, the results based on studies of the motion of planetesimals at initial data considered by me should not differ much for the models with and without consideration of the star binary. Such consideration may be important for some other problems (e.g. for studies of the motion of planetesimals that left the Hill sphere of the star).

The feeding zone of a planet is the zone from which planetesimals could collide a growing planet (Safronov, 1972). For the feeding zone of Proxima Centauri $c$, Ipatov (2023) obtained that $a_c-a_{min002}=0.04a_c+0.02a_{min002}+2.54a_c\cdot\mu^{1/3}$, $a_{max002}-a_c=0.04a_c+0.02a_{max002}+2.40a_c\cdot\mu^{1/3}$ at $e_o=0.02$ and $a_c-a_{min015}=0.04a_c+0.15a_{min015}+2.23a_c\cdot\mu^{1/3}$, $a_{max015}-a_c=0.04a_c+0.15a_{max015}+4.3a_c\cdot\mu^{1/3}$ at $e_o=0.15$, where $a_c$ and $e_c=0.04$ are the semi-major axis and eccentricity of the orbit of planet $c$, $\mu$ is the ratio of the mass of planet $c$ to the mass of the star, $a_{min002}$, $a_{max002}$, $a_{min015}$ and $a_{max015}$ are the minimum and maximum initial values of the semi-major axes of orbits of planetesimals for the feeding zone of planet $c$ at initial eccentricities $e_o$ of the planetesimal orbits equaled to 0.02 and 0.15, respectively. However, there can be some remaining planetesimals moving in resonant orbits inside the above feeding zone and empty resonant zones outside this feeding zone.

**The series *MP* and *F* of calculations with estimates of the frequency of collisions of planetesimals with planets based on the arrays of orbital elements of planetesimals**

Earlier in abstracts (Ipatov, 2021a), other studies of the evolution of orbits of planetesimals in the Proxima Centauri system were presented with initial data similar to those discussed in the previous subsection. In those calculations, planets and planetesimals were considered as point masses (material points) during integration. In contrast to the *C* calculations, the moments of collisions of planetesimals with planets were not calculated, and planetesimals that could collide with planets were not excluded from integrations. Such series of calculations below are denoted as the *MP* and *F* calculations. Initial data for **the series *MP*** were the same as those for the *C* calculations discussed above. The results obtained for the series *MP* are used below only for discussion on the probabilities of collisions of planetesimals with planets $b$ and $d$. In older **series *F*** a greater mass of planet $c$ ($m_c=12m_E$, according to Kervella et al., 2020) and other values of eccentricity of its orbit were considered. For the *F* calculations, the initial eccentricity $e_c$ of the orbit of exoplanet $c$ equaled to 0 or 0.1, $i_c=e_c/2$, and $e_b=i_b=0$. In all series for calculations of radii of planets, the densities of Proxima Centauri $b$ and $c$ equaled to the densities of the Earth and Uranus, respectively.

For the *MP* and *F* series of calculations, the probabilities of collisions of planetesimals with exoplanets were calculated based on the obtained arrays of orbital elements of migrated planetesimals stored with a step of 100 yr. Similar approach was used in (Ipatov and Mather, 2003, 2004a,b; Marov and Ipatov, 2018; Ipatov, 2020, 2021b) for studies of the delivery of bodies from the zone of the giant planets and the outer asteroid belt to the terrestrial planets. Other (than those for the Solar System) masses and radii of the star and exoplanets were considered in the calculations of the probabilities for the Proxima Centauri planetary system. For each set of the orbital elements of planetesimals, at time $t$ the probabilities of collisions of planetesimals with an exoplanet were calculated for a time interval $d_t=100$ yr. These probabilities were summed up over all sets of orbital elements for different times.

In our previous calculations of the orbital evolution of bodies in the Solar System, the calculated probability of a collision of a body with a planet was always less than 1. In the case of calculations of probabilities of collisions of planetesimals with considered exoplanets in the *MP* and *F* series, for a few planetesimals the calculated sum of the probabilities of a collision of $i$th planetesimal with an exoplanet at some times could exceed 1. Therefore, the previous algorithm of calculations of the probabilities was corrected so that the calculated probability of a collision of $i$th planetesimal with an exoplanet could not exceed 1. If the probability of a collision of a planetesimal with an exoplanet reached 1 with time, then for a later time this planetesimal was not considered for calculation of the probability. Without exclusion of the planetesimal which probability reached 1 in the considered calculations, for a small fraction of planetesimals the probability of a collision of a planetesimal with planet $c$ could reach several tens and exceed 100 whereas such probabilities for most other planetesimals were much smaller and smaller than 1. In each calculation variant, the probability of a collision of a planetesimal with an exoplanet was calculated as the sum of probabilities for 250 planetesimals divided by 250. Probabilities of collisions of some planetesimals with an exoplanet could be zero.

In calculations of the probability $p_{dts}$ of a close encounter of a planetesimal with an exoplanet to the distance equal to the radius $r_s$ of the considered sphere (the sphere of action of the exoplanet) for the time interval $d_t$, the following formulas were used in the 3D model (Ipatov, 1988, 2019): $p_{dts}=d_t/T_3$, where $T_3=2\pi^2 \cdot k_p \cdot k_v \cdot T_s \cdot R^2 \cdot \Delta i/(r_s^2 \cdot k_{fi})$ is the characteristic time elapsed before the encounter. Here $\Delta i$ is the angle (expressed in radians) between the orbital planes of the encountering celestial objects, $R$ is the distance from the encounter location to the star, $k_{fi}$ is the sum of angles (expressed in radians) with vertices in the star, within which the distance between the projections of orbits onto the plane of the ecliptic is smaller than $r_s=R \cdot \mu^{0.4}$, $\mu$ is the ratio of the sum of masses of encountering objects to the stars's mass, $T_s$ is the synodic period, $k_p=P_2/P_1$, $P_2>P_1$, $P_i$ is the rotation period of the $i$th object (a planetesimal or an exoplanet) about the star, $k_v=(2a/R-1)^{1/2}$, and $a$ is the semi-major axis of the planetesimal's orbit. The coefficient $k_v$ was added by Ipatov and Mather (2004a) in order to take into account the dependence of the encounter velocity on the planetesimal's position on the eccentrical orbit.

The collision probability for two celestial objects, which entered the sphere of action (the Tisserand sphere), was assumed to be $p_{dtc}=(r_\Sigma/r_s)^2 \cdot (1+(v_{par}/v_{rel})^2)$, where $v_{par}=(2Gm_\Sigma/r_\Sigma)^{1/2}$ is the parabolic velocity, $v_{rel}$ is the relative velocity of the objects coming to the distance $r_s$ of each other, $r_\Sigma$ is the sum of the radii of encountering objects of a total mass $m_\Sigma$, and $G$ is the gravitational constant. The probability $p_{dt}$ of the collision of a planetesimal with an exoplanet during time $d_t$ is $p_{dts} \cdot p_{dtc}$. The values of $p_{dt}$ were summed up through the considered time interval $T$ or the whole dynamical lifetime of a planetesimal (if it was less than $T$).

At $\Delta i < r_s^* = r_s/R$, the formula for $T_3$ overestimates the probabilities $p_{dts}$ of encounters of bodies up to a sphere radius $r_s$ by about a factor of $r_s^*/\Delta i$. On the other hand at $\Delta i < r_s^*$, the value of $p_{dtc}$ is underestimated (up to a factor of $r_s/r_\Sigma$ at small ratio $r_\Sigma/r_s$). So at small $\Delta i$ the probability $p_{dt}=p_{dts} \cdot p_{dtc}$ of a collision is not overestimated as much as $p_{dts}$. For the considered sphere of action of planet $c$ we have $r_s^*=0.03$. For comparison, initial inclinations of planetesimals equaled to 0.014 and 0.106 rad at $e_o$ equal to 0.02 or 0.15, respectively. Below in the paper it is shown that the probabilities of collisions of planetesimal with planets $b$ and $c$ obtained in the $C$ integration are even greater than the values of the probabilities that are based on the above analytical formulas and on the arrays of orbital elements obtained in the $C$ integrations. The obtained results do not show large differences in the probabilities of collisions of planetesimals with planets for the model $C$ (which considers collisions of planetesimals with planets during integration of motion equations) and the model $MP$ (which uses formulas for calculations of the probabilities of collisions of planetesimals with planets).

In contrast to the $C$ series, the $MP$ and $F$ calculations allow one to estimate the probabilities of collisions of planetesimals with exoplanet $d$ (with $a_d=0.02895$ AU, $m_d=0.29m_E$, $e_d=i_d=0$; Suarez´ Mascareno et al., 2020), though this planet was not included in the integration of motion equations. Probabilities $p_b^*$, $p_c^*$, and $p_d^*$ of collisions of a planetesimal with exoplanets $b$, $c$, and $d$, respectively, were calculated for several time intervals $T$ (equaled to 0.1, 1, 10, 20, and 50 Myr). Here and below the values of the probabilities of collisions calculated based on the arrays of orbital elements of migrated planetesimals are marked by *. For considered time intervals, the fraction of planetesimals ejected into hyperbolic orbits $p_{ej}=N_{ej}/250$ and the fraction of planetesimals collided with the star $p_{st}=N_{st}/250$ were calculated while integration of the motion of planetesimals (the radius of the star was considered to be equal to 0.154 of that of the Sun). For each variant with 250 initial planetesimals, $N_{ej}$ is the number of ejected planetesimals, and $N_{st}$ is the number of planetesimals collided with the star. The values of probabilities $p_{ej}$, $p_{st}$, $p_b^*$, $p_c^*$, and $p_d^*$ were calculated based of the results of numerical simulation of the motion of planetesimals, which did not include collisions of planetesimals with exoplanets. Based on the above probabilities, the models $A$ and $B$ (see Appendix A) correct the estimates of probabilities of collisions of planetesimals with planets estimating in different ways the decrease in the number of migrated planetesimals both due to ejections into hyperbolic orbits and collisions with exoplanets. The intervals in which the probabilities are located are discussed in Appendix B.

The approach that is similar to the *MP* calculations allows one to estimate the probabilities of collisions of bodies-planetesimals with planets when such probabilities are small and the number of considered bodies is not large. For example, for calculations that simulate direct collisions of small bodies with a planet, it is needed to consider many millions of bodies if the probability of a collision of a body with a planet is about $10^{-5}$. Before beginning calculations, I supposed that the probabilities of collisions of planetesimals with planets *b* and *d* could be small, so the calculations began with the *MP* approach.

## RESULTS OF CALCULATIONS

**The probabilities of collisions of planetesimals with Proxima Centauri *c* and the probabilities of their ejection into hyperbolic orbits**

*Calculations at the present mass of planet c*

The fraction $p_{el}$ of planetesimals that still moved in elliptical orbits, the probability $p_c$ of a collision of a planetesimal with Proxima Centauri *c*, the probability $p_{ej}$ of ejection of a planetesimal, and the ratio $p_{cej}=p_c/p_{ej}$ vs. $a=a_{min}+0.05$ AU are presented at several time intervals for the *C* calculations in Figs. 1, 2, 3, and 4, respectively. For data presented in the figures, planetesimals considered to be ejected into hyperbolic orbits when their distances from the star reached 500 AU. Each point on the figures corresponds to the mean value of a probability for 250 planetesimals. The evolution of orbits of planetesimals can be chaotic because of their close encounters with exoplanets. However, the obtained probabilities $p_{el}$, $p_c$, $p_{ej}$, and $p_c/p_{ej}$ didn't differ much at integrations with a different time step $t_s$ (equaled to 1, 0.1, 0.2, 0.5, or 2 days). These probabilities can differ much for different $a_{min}$. For the values of $a_{min}$ and $e_o$ presented in Table 2, the evolution finished in less than 40 Myr. The main changes in $p_{el}$, $p_c$, and $p_{ej}$ were typically during the first 10 Myr. After 10 Myr, the value of $p_{el}$ changed usually by less than 0.1-0.2. Only at $a_{min}=2.1$ AU and $e_o=0.15$, the main changes in $p_{el}$ and $p_{ej}$ were after 10 Myr. The planetesimals that moved in elliptical orbits at $t=100$ Myr usually still continued to move in such orbits at least for hundreds of millions of years.

Table 2. The times of depletion of disks (at $t_s=1^d$) that are less than 100 Myr in the *C* variants. Distances of all ejected planetesimals from the star reached 500 and 1200 AU in $T_{500}$ and $T_{1200}$ Myr, respectively. Other planetesimals collided with planets or the star.

| $a_{min}$ (AU) | 1.3 | 1.6 | 1.3 | 1.4 | 1.6 |
|---|---|---|---|---|---|
| $e_o$ | 0.02 | 0.02 | 0.15 | 0.15 | 0.15 |
| $T_{500}$ (Myr) | 30.34 | 35.19 | 37.71 | 89.45 | 39.66 |
| $T_{1200}$ (Myr) | 30.36 | 35.58 | 37.82 | 89.50 | 39.66 |

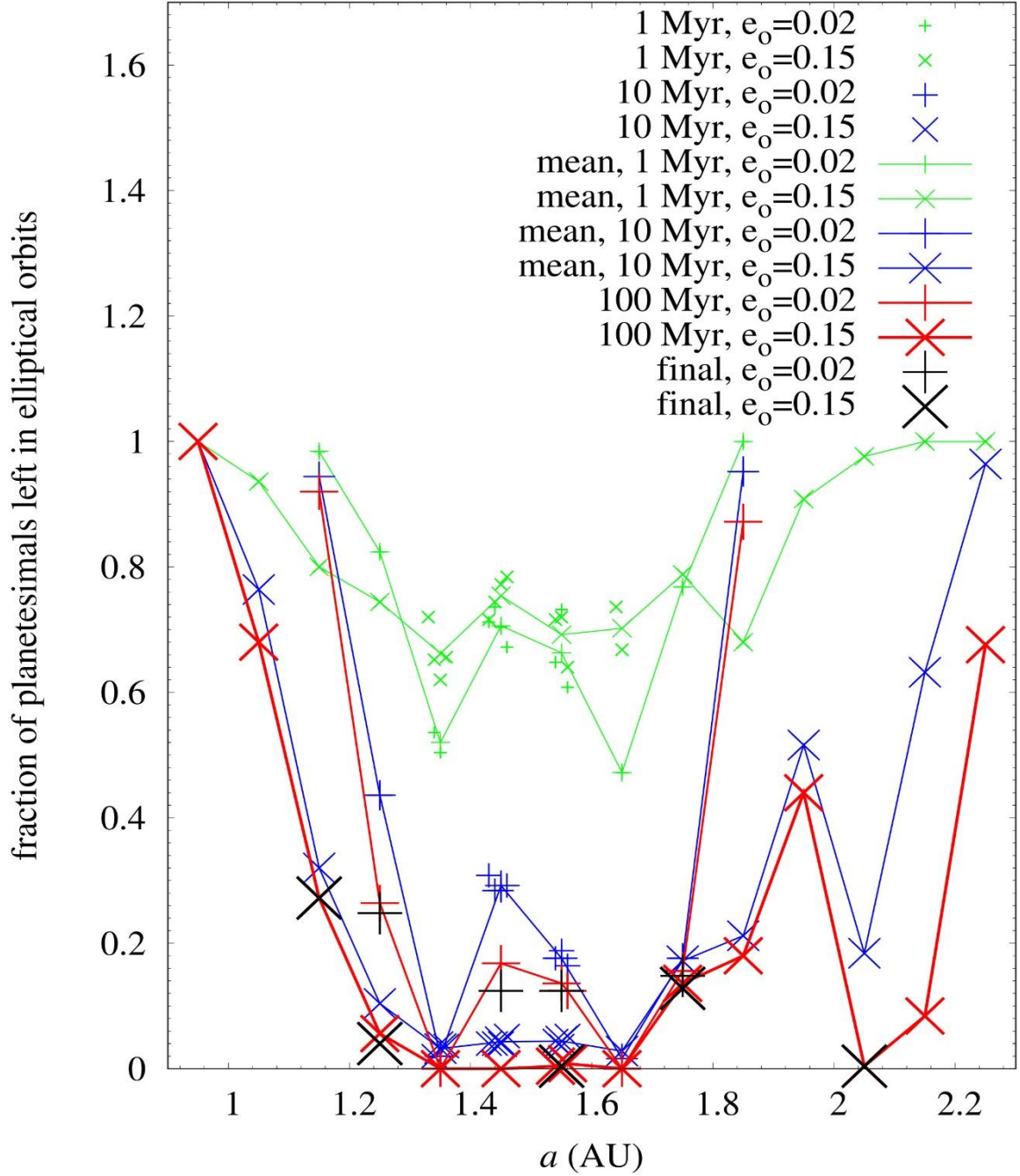

Fig. 1. The fraction of planetesimals that left in elliptical orbits vs. $a=a_{min}+0.05+a_{ts}$ AU at several time intervals (equal to 0.1, 1, 10, 100 Myr, and at the final time usually equaled to 1 Gyr) for the $C$ variants. Initial semi-major axes of orbits of planetesimals were between $a_{min}$ and $a_{min}+0.1$ AU, and initial eccentricities $e_o$ were equal to 0.02 or 0.15. $a_{ts}$ is equal to 0, -0.03, -0.02, -0.01, and 0.01 AU at an integration time step $t_s$ equal to $1^d$, $0.1^d$, $0.2^d$, $0.5^d$, and $2^d$, respectively. Each sign on the figure corresponds to the mean value for 250 planetesimals. The sign that marked as "mean" in the figure corresponds to the mean value of probabilities obtained at different $t_s$, which can be based on calculations with about 1000 planetesimals.

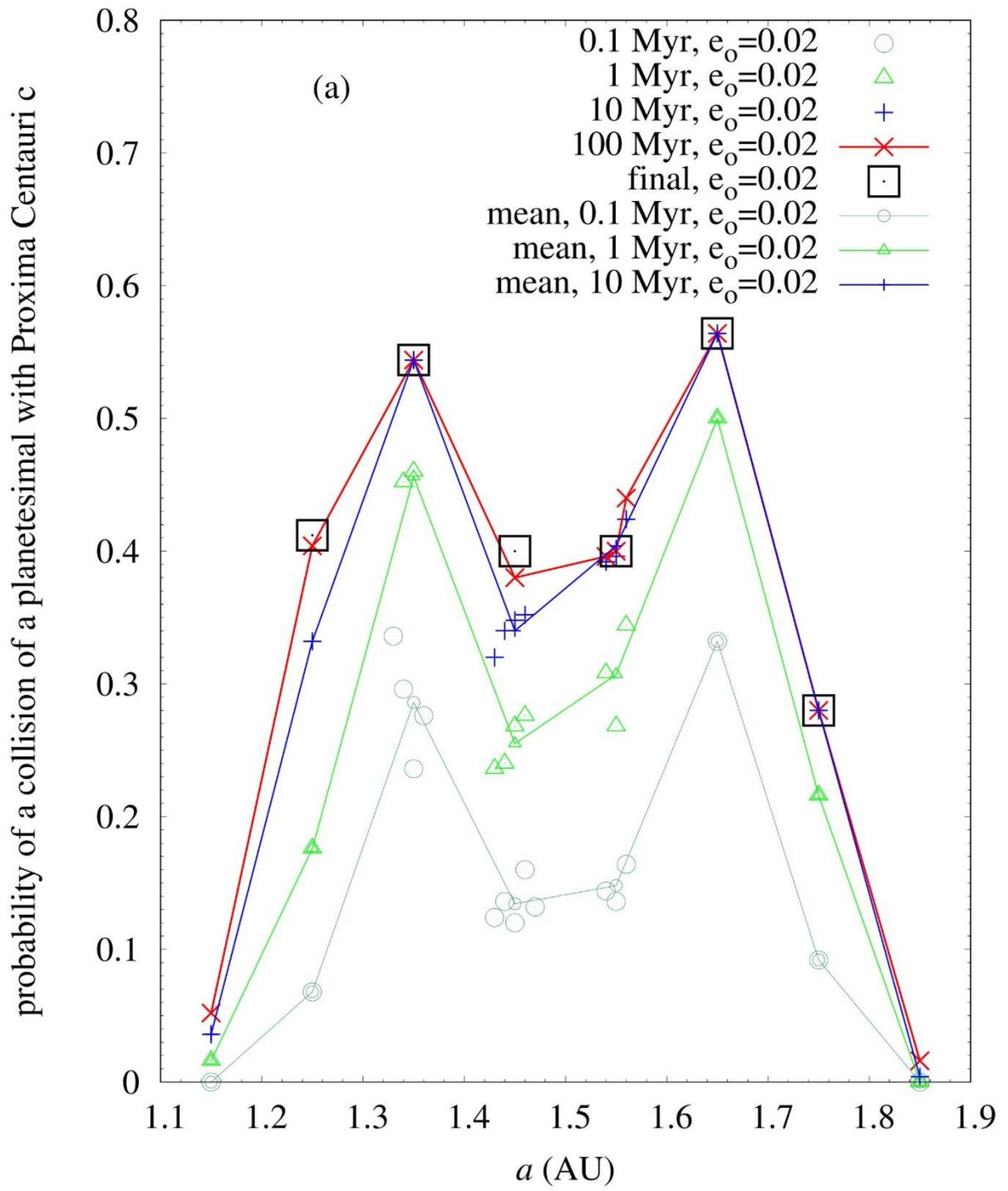

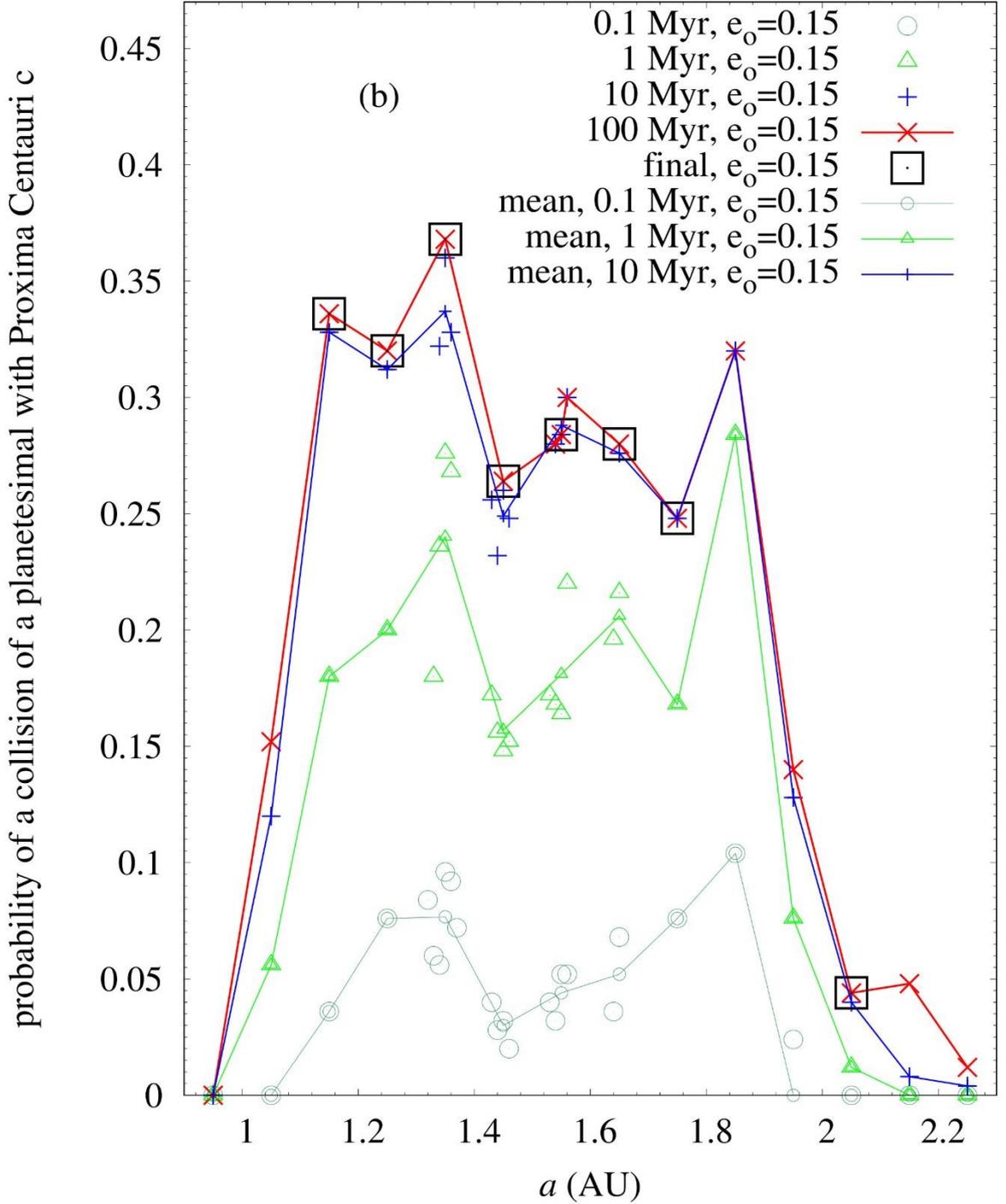

Fig. 2. The probability $p_c$ of a collision of a planetesimal with Proxima Centauri $c$ vs. $a=a_{min}+0.05+a_{ts}$ AU at several time intervals for the $C$ variants. Initial semi-major axes of orbits of planetesimals were between $a_{min}$ and $a_{min}+0.1$ AU, and initial eccentricities $e_o$ were equal to 0.02 (Fig. 2a) or 0.15 (Fig. 2b). $a_{ts}$ is equal to 0, -0.03, -0.02, -0.01, and 0.01 AU at an integration time step $t_s$ equal to $1^d$, $0.1^d$, $0.2^d$, $0.5^d$, and $2^d$, respectively. Each sign on the figure corresponds to the mean value for 250 planetesimals. The sign that marked as "mean" in the figure corresponds to the mean value of probabilities obtained at different $t_s$, which can be based on calculations with about 1000 planetesimals.

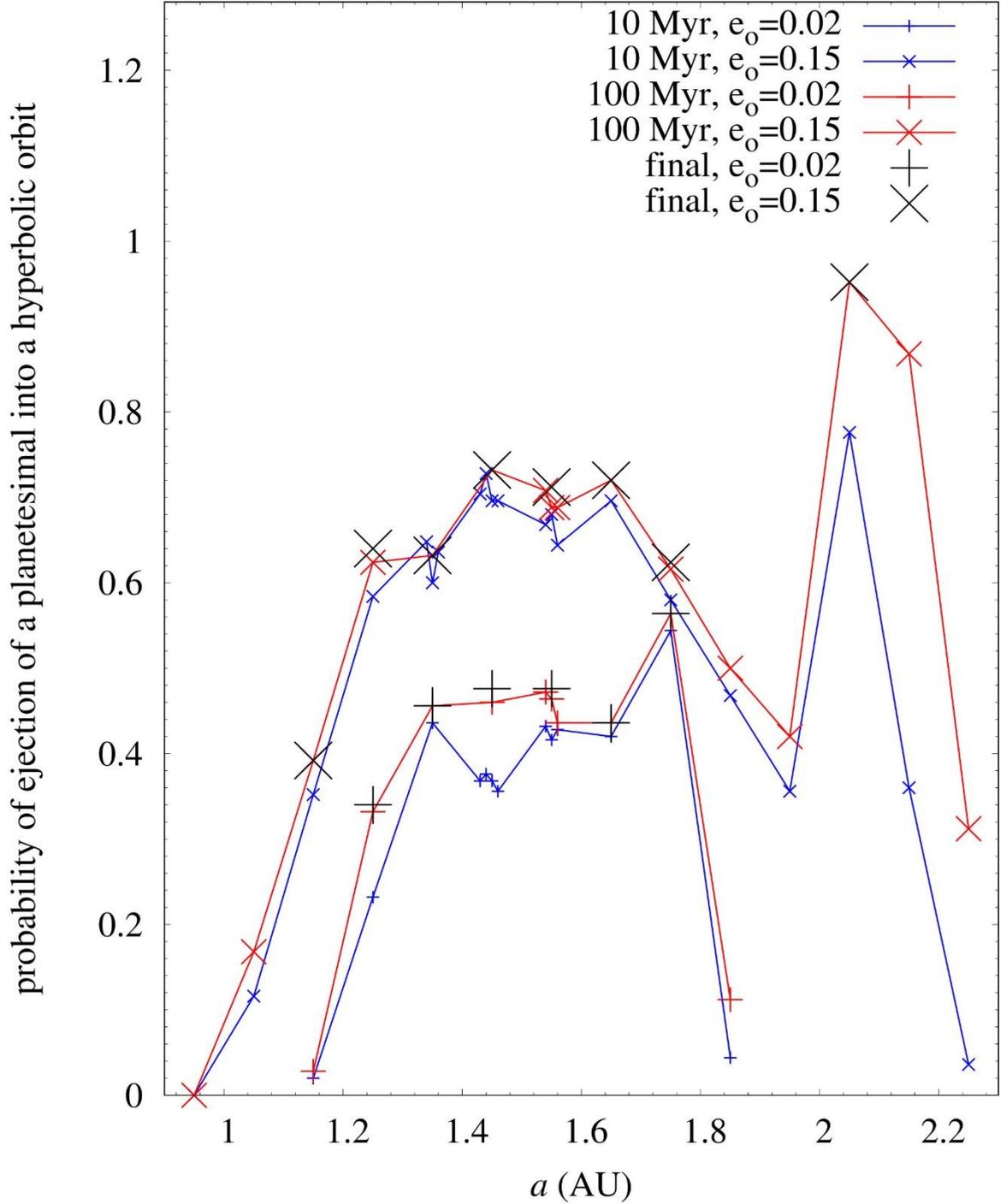

Fig. 3. The probability $p_{ej}$ of an ejection of a planetesimal into a hyperbolic orbit vs. $a=a_{min}+0.05+a_{ts}$ AU at several time intervals for the $C$ variants. Initial semi-major axes of orbits of planetesimals were between $a_{min}$ and $a_{min}+0.1$ AU, and initial eccentricities $e_o$ were equal to 0.02 or 0.15. $a_{ts}$ is equal to 0, -0.03, -0.02, -0.01, and 0.01 AU at an integration time step $t_s$ equal to $1^d$, $0.1^d$, $0.2^d$, $0.5^d$, and $2^d$, respectively. Each sign on the figure corresponds to the mean value for 250 planetesimals.

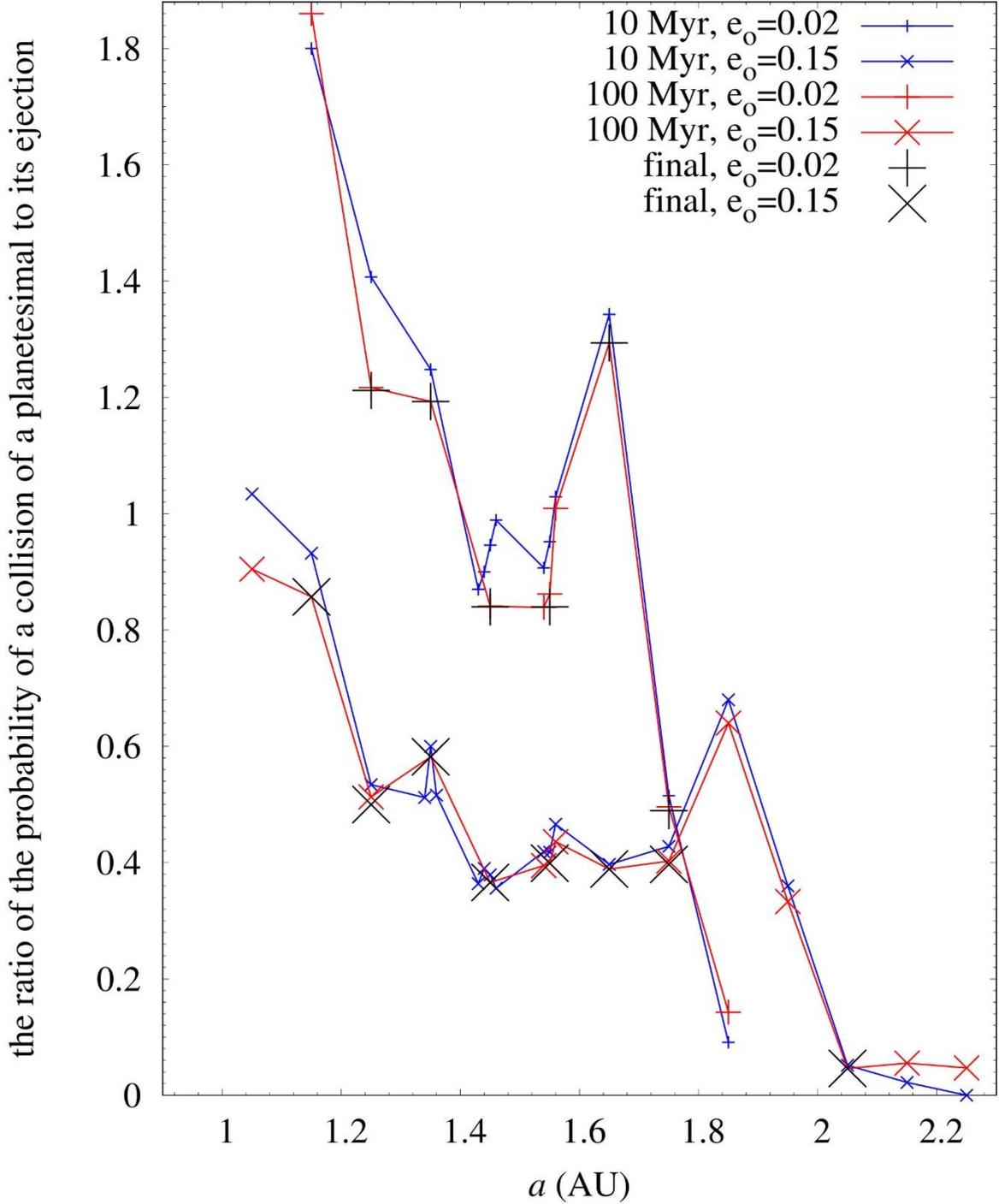

Fig. 4. The ratio $p_{cej}$ of the probability $p_c$ of a collision of a planetesimal with Proxima Centauri $c$ to the probability of its ejection into a hyperbolic orbit vs. $a=a_{min}+0.05+a_{ts}$ AU at several time intervals for the $C$ variants. Initial semi-major axes of orbits of planetesimals were between $a_{min}$ and $a_{min}+0.1$ AU, and initial eccentricities $e_o$ were equal to 0.02 or 0.15. $a_{ts}$ is equal to 0, -0.03, -0.02, -0.01, and 0.01 AU at an integration time step $t_s$ equal to $1^d$, $0.1^d$, $0.2^d$, $0.5^d$, and $2^d$, respectively. Each sign on the figure corresponds to the mean value for 250 planetesimals.

The role of collisions with the star and planet $b$ in the decrease in the number of planetesimals in a disk was small (see subsection "Probabilities of collisions of planetesimals with Proxima Centauri $b$ and $d$ and the star"). For $e_o$=0.02, the final values of $p_c$ were maximum (about 0.55) at $a_{min}$=1.3 AU and $a_{min}$=1.6 AU and exceeded 0.4 at 1.2≤$a_{min}$≤1.6 AU. For $e_o$=0.15, the final values of $p_c$ were typically smaller than at $e_o$=0.02 and were maximum (about 0.37) at $a_{min}$=1.3 AU and exceeded 0.25 at 1.2≤$a_{min}$≤1.8 AU. The ratio $p_{cej}$=$p_c/p_{ej}$ of the probability $p_c$ of a collision of a planetesimal with Proxima Centauri $c$ to the probability $p_{ej}$ of its ejection into a hyperbolic orbit was about 0.84 at 1.4≤$a_{min}$≤1.5 AU, and it was in the range (1.2, 1.3) at $a_{min}$ equal to 1.2, 1.3, and 1.6 AU at $e_o$=0.02. For $e_o$=0.15 and 1.2≤$a_{min}$≤1.7 AU, the ratio $p_c/p_{ej}$ was about 0.4-0.6, i.e. on average it was about twice smaller than for $e_o$=0.02 and 1.2≤$a_{min}$≤1.6 AU. At 2≤$a_{min}$≤2.2 AU and $e_o$=0.15, the values of $p_c/p_{ej}$ were not more than 0.05. Such planetesimals changed the orbit of planet $c$, but almost didn't increase its mass.

All the above estimates were for the model $C$, in which planetesimals that collided with planets were excluded from integration. It is discussed in Appendix A, how it is possible to estimate the probability $p_c^{\#}$ of a collision of a planetesimal with planet $c$ and the probability $p_{ej}^{\#}$ of an ejection of a planetesimal into a parabolic-hyperbolic orbit for the model $MP$, in which planets and planetesimals are considered as point masses in integration. The probabilities of collisions of planetesimals with planet $c$ for the $MP$ series of calculations are discussed in Appendix B. The ratio $p_c/p_c^{\#}$ of the probabilities of collisions of planetesimals with planet $c$ for the $C$ calculations to that for the $MP$ calculations was about 1.6-2. The mean value of the ratio of the probabilities of collisions of planetesimals with planet $c$ to the probabilities of their ejection into hyperbolic orbits was smaller by a factor of 2-3 for the $MP$ series than for the $C$ calculations. It testifies in favor of that the $MP$ calculations underestimated collision probabilities of planetesimals with planet $c$.

*Calculations at a smaller mass of planet c*

The values of the probabilities $p_{el}$, $p_c$, $p_{ej}$, and $p_c/p_{ej}$ presented in Figs. 1-4 and discussed above were calculated for the present mass $m_c$=7$m_E$ of planet $c$. In order to estimate how such probabilities depend on the mass of the planet, the $C$ calculations were also made for smaller values of the mass of a planet, which had the same orbit as planet $c$. The ratio $k_c$=$m_e/m_c$ of the mass of the planet to the mass of planet $c$ was considered to be equal to 0.5 or 0.1. The results of calculations are presented in Table 3 for $a_{min}$ equal to 1.4 and 1.5 AU, $e_o$=0.02 and $e_o$=0.15. At $e_o$=0.02 and $k_c$=0.1, $p_c$ was about 0.1 already at $T$=0.1 Myr, i.e. for this case a planet could accumulate about 10 per cent of planetesimals in 0.1 Myr and increased its mass. At a greater $e_o$, the growth of a planet mass was smaller. Probably, calculations with small eccentricities $e_o$ could better correspond to smaller values of the mass of a growing planet (e.g. at $k_c$=0.1), and mean eccentricities of orbits of planetesimals were not small at almost present mass of planet $c$ (at $k_c$=1).

At $T$=0.1 Myr, the values of $p_c$ and $p_{ej}$ were usually smaller for smaller $k_c$. However, at $T$=100 Myr in most variants the values of $p_c$ were greater for smaller $k_c$. The greater fraction of planetesimals collided with a planet with a smaller mass was because the ejection into hyperbolic orbits was smaller for smaller masses, and the planet accumulated with time most of planetesimals which were not far from its orbit. In Table 3 the ratio $p_c/p_{ej}$ was greater at $k_c$=0.5 than at $k_c$=1.0 by a factor from 2.2 to 3.0 at $T$=10 Myr and by a factor from 1.2 to 1.8 at $T$=100 Myr. For 10≤$T$≤100 Myr the value of $p_c$ was greater at $k_c$=0.5 than at $k_c$=1 by a factor from 0.96 to 1.3. The ejection of planetesimals into hyperbolic orbits was much less at $k_c$=0.1 than at $k_c$=1.

Table 3. The fractions $p_c$ and $p_{ej}$ of planetesimals collided with growing planet $c$ and ejected into hyperbolic orbits, respectively, at several values of $a_{min}$, $e_o$, $T$, and the ratio $k_c=m_e/m_c$ of the mass of the growing planet to the present mass of planet $c$ for the series $C$.

| $a_{min}$ (AU) | $e_o$ | $T$ (Myr) | $k_c=0.1$ $p_c$ | $k_c=0.1$ $p_{ej}$ | $k_c=0.1$ $p_c/p_{ej}$ | $k_c=0.5$ $p_c$ | $k_c=0.5$ $p_{ej}$ | $k_c=0.5$ $p_c/p_{ej}$ | $k_c=1$ $p_c$ | $k_c=1$ $p_{ej}$ | $k_c=1$ $p_c/p_{ej}$ |
|---|---|---|---|---|---|---|---|---|---|---|---|
| 1.4 | 0.02 | 0.1 | 0.096 | 0 | | 0.08 | 0 | | 0.12 | 0 | |
| | | 1 | 0.208 | 0 | | 0.2 | 0 | | 0.268 | 0.028 | 9.57 |
| | | 10 | 0.376 | 0 | | 0.34 | 0.124 | 2.74 | 0.348 | 0.368 | 0.95 |
| | | 100 | 0.548 | 0.036 | 15.2 | 0.364 | 0.284 | 1.28 | 0.38 | 0.368 | 1.03 |
| 1.5 | 0.02 | 0.1 | 0.08 | 0 | | 0.164 | 0 | | 0.136 | 0 | |
| | | 1 | 0.232 | 0 | | 0.336 | 0 | | 0.328 | 0.048 | 6.83 |
| | | 10 | 0.448 | 0 | | 0.484 | 0.168 | 2.88 | 0.396 | 0.416 | 0.95 |
| | | 100 | 0.672 | 0.068 | 9.88 | 0.52 | 0.344 | 1.51 | 0.4 | 0.472 | 0.85 |
| 1.4 | 0.15 | 0.1 | 0.008 | 0 | | 0.024 | 0 | | 0.032 | 0 | |
| | | 1 | 0.048 | 0 | | 0.104 | 0 | | 0.148 | 0.08 | 1.85 |
| | | 10 | 0.252 | 0.004 | 63 | 0.276 | 0.312 | 0.88 | 0.26 | 0.7 | 0.37 |
| | | 100 | 0.456 | 0.14 | 3.26 | 0.32 | 0.508 | 0.63 | 0.268 | 0.732 | 0.37 |
| 1.5 | 0.15 | 0.1 | 0.016 | 0 | | 0.044 | 0 | | 0.052 | 0 | |
| | | 1 | 0.052 | 0 | | 0.14 | 0.004 | 35 | 0.164 | 0.116 | 1.41 |
| | | 10 | 0.204 | 0 | | 0.292 | 0.316 | 0.92 | 0.284 | 0.68 | 0.42 |
| | | 100 | 0.468 | 0.22 | 2.13 | 0.316 | 0.664 | 0.48 | 0.284 | 0.712 | 0.40 |

**Probabilities of collisions of planetesimals with Proxima Centauri $b$ and $d$ and the star**

*Probabilities of collisions of planetesimals with planet b and the star for the C integrations*

The planetesimals that came from beyond the water-ice condensation line (e.g. from the feeding zone of Proxima Centauri $c$) could deliver water and volatiles to inner exoplanets. Such studies are interesting for understanding the possible habitability of the inner exoplanets. If averaged over many planetesimals, the probabilities of collisions of a planetesimal with Proxima Centauri $b$ and $d$ were much smaller than the probability of a collision of a planetesimal with Proxima Centauri $c$.

For the series $C$, collisions with planet $b$ or with the star were obtained only in calculations at $e_o$=0.15 with $a_{min}$ equal to 1.4 or 1.5 AU. The times $t_{col}$ (in Myr) elapsed before such collisions with planet $b$ or the star are presented in Table 4. Such collisions were obtained in calculations with different time steps $t_s$ of integration and at different masses of growing planet $c$. The times $t_{col}$ were between 1 and 4 Myr at the present mass $m_c$ of planet $c$, and they were about 40-50 Myr for the growing planet $c$ with masses equal to $0.5m_c$ or $0.1m_c$.

Table 4. The times $t_{col}$ (in Myr) elapsed before collisions of planetesimals with planet $b$ or the star for the $C$ calculation variants at $e_o$=0.15, $a_{min}$, a time step of integration $t_s$, and the ratio $k_c=m_e/m_c$ of the mass of a growing planet $c$ to the present mass of planet $c$.

| $a_{min}$ (AU) | 1.4 | 1.5 | 1.5 | 1.4 | 1.5 | 1.5 | 1.4 | 1.5 |
|---|---|---|---|---|---|---|---|---|
| $t_s$ (days) | 1 | 0.5 | 2 | 2 | 0.5 | 1 | 1 | 1 |
| $k_c$ | 1 | 1 | 1 | 1 | 1 | 1 | 0.5 | 0.1 |
| $t_{col}$ (Myr) | 2.20 | 3.15 | 2.69 | 4.07 | 1.16 | 1.64 | 50.92 | 41.51 |
| Collision with | $b$ | $b$ | $b$ | star | star | star | star | $b$ |

For calculations with $e_o$=0.15, $k_c$=1 (i.e., $m_c=7m_E$), only three planetesimals collided with planet $b$. For all of them $a_{min}$ equaled to 1.4 or 1.5 AU. Variants with such $a_{min}$ and considered time interval $T\geq100$ Myr (including the variants for which the evolution ended at smaller times than 100 Myr) included 1000 planetesimals, and those variants with $T\geq10$ Myr included 1500 planetesimals. For $e_o$=0.15, $k_c$=1, $T≥10$ Myr, there were 3000 planetesimals in all considered variants with $1.2≤a_{min}≤1.7$ AU and 4500 planetesimals with $1.1≤a_{min}≤2.2$ AU (for such values of $a_{min}$, less than a half of initial planetesimals were left in elliptical orbits at the end of integration). Therefore, the probability $p_b$ of a collision of a planetesimal with planet $b$ was about $10^{-3}$ at $e_o$=0.15. The value of the probability $p_{star}$ of a collision of a planetesimal with the star was about the same as $p_b$.

For calculations with $e_o$=0.02, $1.2≤a_{min}≤1.7$ AU, $k_c$=1, $T≥10$ Myr (in these calculations less than a half of initial planetesimals left in elliptical orbits), 2500 initial planetesimals were considered, but collisions with planet $b$ were not obtained. So the probability of a collision of a planetesimal with planet $b$ could be not more than $4·10^{-4}$ at $e_o$=0.02.

At $1.4≤a_{min}≤1.5$ AU and $e_o$=0.15, there was one collision of a planetesimal with planet $b$ among 250 planetesimals at $k_c$=0.1, and one collision with the star among 500 planetesimals at $k_c$=0.5. For comparison, at $k_c$=1 for 1500 planetesimals, three planetesimals collided with planet $b$ (i.e., $p_b$=$2·10^{-3}$) and two planetesimals collided with the star. So at $e_o$=0.15 the probability of a collision of a planetestimal from the feeding zone of planet $c$ with growing planet $b$ (or the star) could be of the same order of magnitude during the growth of mass of planet $c$ from $0.1m_c$ to its present mass $m_c$. Such estimate is for the model for which the growing planet moved in the same orbit as planet $c$, and the mass and orbit of planet $b$ had present values.

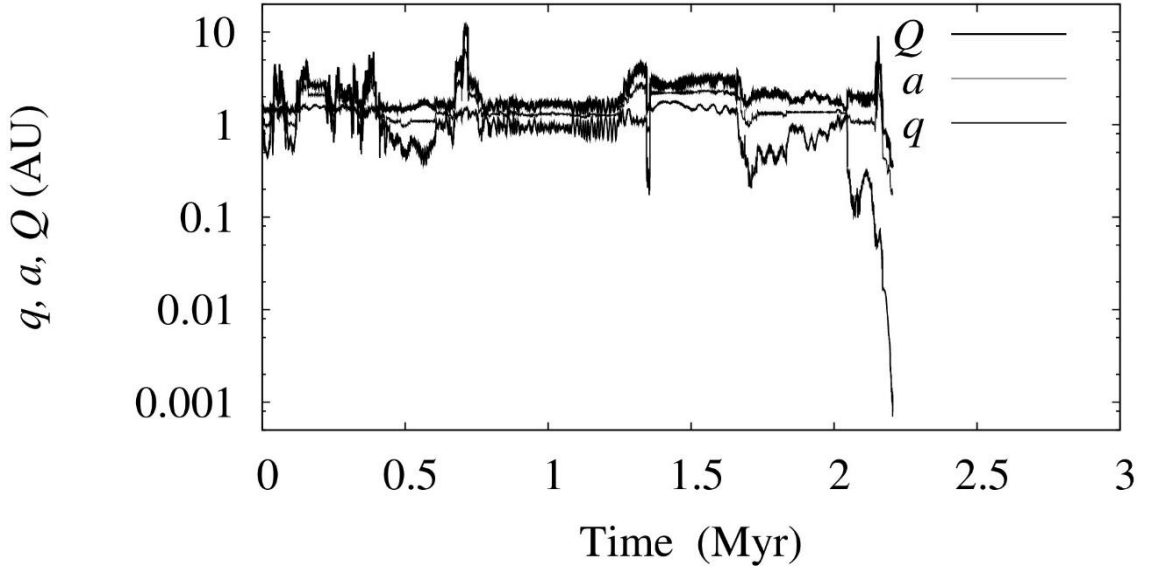

Fig. 5. Time evolution of a semi-major axis $a$, perihelion and aphelion distances $q$ and $Q$ of an orbit of a planetesimal (at $a_o$=1.41566 AU) for the $C$ calculation variant at $a_{min}$=1.4 AU, $e_o$=0.15, $k_c$=1, and $t_s$=$1^d$.

The time evolution of semi-major axes $a$, perihelion and aphelion distances $q$ and $Q$ of orbits of two planetesimals that collided with planet $b$ is presented in Figs. 5-6 for the $C$ calculation variants at $e_o$=0.15 and $k_c$=1. Such collisions were obtained in calculations with different integration time steps $t_s$. Figs. 5 and 6 are for $t_s$=$1^d$ and $0.5^d$, but similar plots were obtained also at $t_s$=$2^d$. Fig. 7 presents the case when a planetesimal collided with planet $b$ for the mass of a growing planet (moving in the same orbit as planet $c$) equal to $0.1m_c$=$0.7m_E$ ($k_c$=0.1). The ratio of the mass of planet $c$ to the mass of the star equals to $1.7×10^{-4}$. This mass ratio allows the growth of eccentricities $e$ of planetesimals to the values up to 1. In most cases at large $e$, the perihelion of an orbit of a planetesimal was close to the orbit of planet $c$. Such planetesimals could reach large distances from the star. For the cases presented in Figs. 5-7, the aphelion of an orbit of a planetesimal was close to the orbit of planet $c$, and its perihelion distance decreased with a growth of $e$.

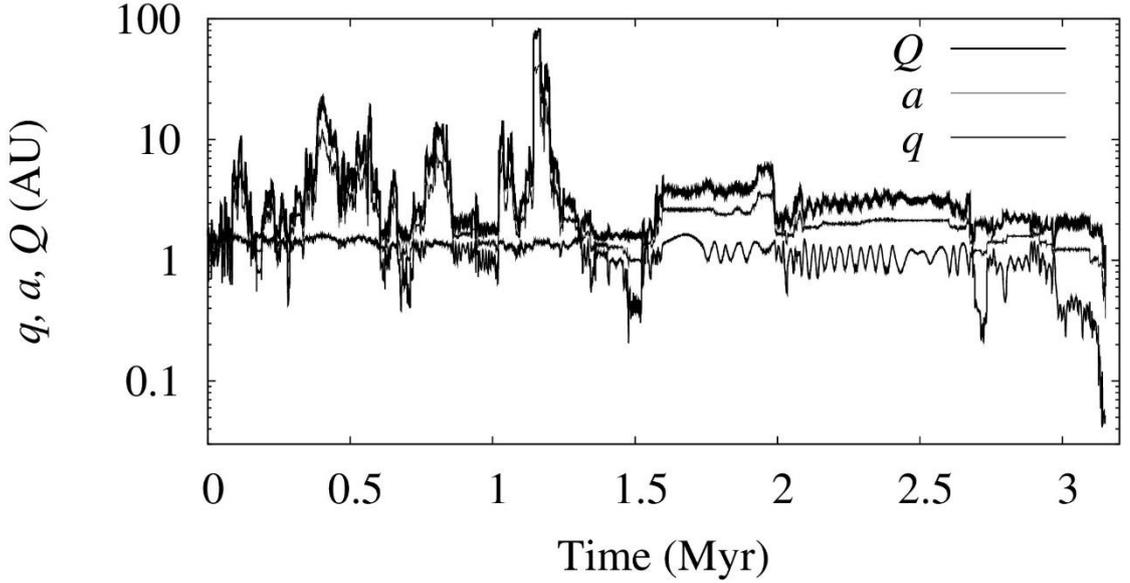

Fig. 6. Time evolution of a semi-major axis $a$, perihelion and aphelion distances $q$ and $Q$ of an orbit of a planetesimal (at $a_o$=1.56315 AU) for the $C$ calculation variant at $a_{min}$=1.5 AU, $e_o$=0.15, $k_c$=1, and $t_s$=0.5$^d$.

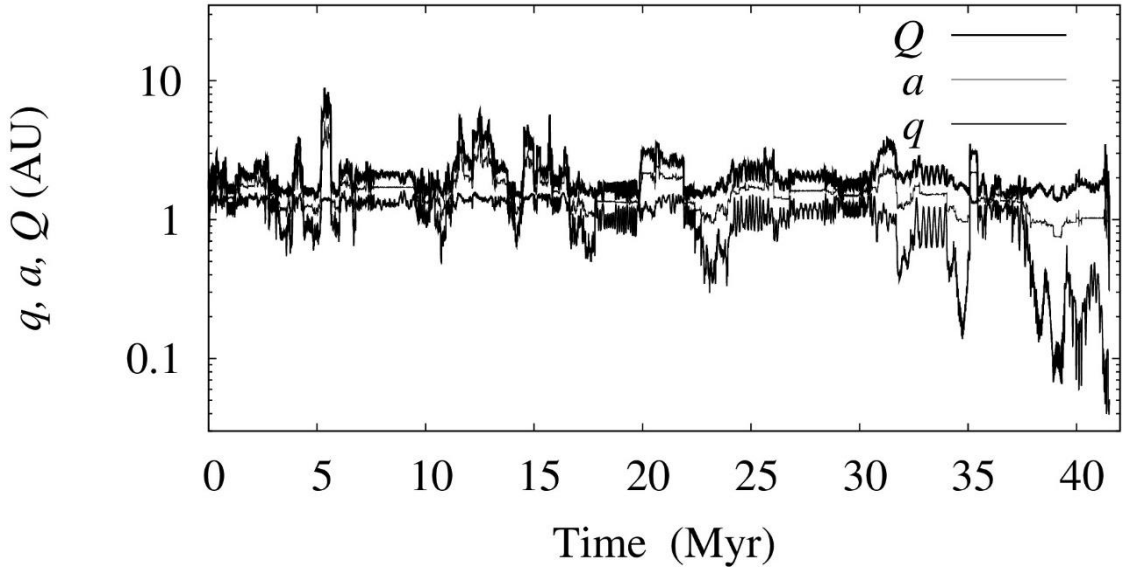

Fig. 7. Time evolution of a semi-major axis $a$, perihelion and aphelion distances $q$ and $Q$ of an orbit of a planetesimal (at $a_o$=1.51481 AU) for the $C$ calculation variant at $a_{min}$=1.5 AU, $e_o$=0.15, $k_c$=0.1, and $t_s$=1$^d$.

*Probabilities of collisions of planetesimals with planets b and d calculated based on arrays of their orbital elements obtained in the series C*

Based on the arrays of orbital elements of migrated planetesimals obtained in the series $C$ of calculations and using the formulas for the probability of a collision of a planetesimal with a planet (similar to calculations for the series $MP$), I calculated the probability $p_c^*$ of a collision of a planetesimal with a planet during considered time interval. The values of $p_c^*$ were typically smaller than the values of $p_c$, which are equal to the ratio of the number of planetesimals collided with planet $c$ to the initial number of planetesimals. For $T$=10 Myr, the ratio $p_c/p_c^*$ was often about 1.5-2.5, i.e. the probabilities of collisions of planetesimals with planet $c$ calculated based on arrays of orbital elements were about twice less than those obtained for the $C$ variants. For the entire considered time interval of evolution of a disk of planetesimals, the ratio often was about 4-6. At $e_o$=0.02 such ratio was even about 10, though mean inclinations were smaller than at $e_o$=0.15.

For the *C* variants presented in Table 4, based on the arrays of orbital elements of the migrated planetesimals that collided with planet *b* or the star, the probabilities $p_b^{*1}$ and $p_d^{*1}$ of collisions of such planetesimal with planets *b* and *d* were calculated. The values of $p_b^{*1}$ were about 0.02-0.06. The values of the ratio $p_b^{*1}/p_d^{*1}$ were between 1 and 4. For such calculations, on average the probability of a collision of a planetesimal with planet *b* was greater than that with planet *d* by about a factor of 2.5. The values of $p_b^*$ and $p_d^*$ averaged over *N* planetesimals equaled to $p_b^{*1}/N$ and $p_d^{*1}/N$, if among these *N* planetesimals there was only one planetesimal with nonzero value of $p_b^{*1}$.

The values of $p_c^*$ and $p_b^{*1}$ that were calculated using some formulas based on the arrays of orbital elements taken from the *C* calculations are smaller (i.e. characteristic times are greater) than the similar probabilities obtained at the *C* integrations. One of possible reasons of this difference may be because actual collisions were often at the beginning of the characteristic times elapsed before the collisions, and for a set of planetesimals, actual collisions were for the planetesimals with smaller times before the collisions, but not for planetesimals with "averaged" times. Ipatov (1993) simulated the evolution of disks of bodies combined at collisions. The mutual gravatitional influence of bodies was taken into account with the method of spheres of action, i.e. the mutual gravitational influence of bodies was calculated only inside the sphere, and bodies moved around the Sun in unperturbed orbits outside the spheres. Orbits changed only during encounters. In the "probabilistic" approach, two encountering bodies were chosen proportional to the probability $p_{ij}$ of their encounter (up to their sphere of action). For a choice of numbers *i* and *j* of two encountering bodies, a random number between 0 and the sum of $p_{ij}$ was simulated. For the "deterministic" method, the matrix of characteristic times $\tau_{ij}$ elapsed before encounters of bodies *i* and *j* was calculated before each encounter, and the smallest $\tau_{ij}$ in this matrix was found. Though the same formulas were used for the calculation of $p_{ij}$ and $\tau_{ij}$, the time of evolution of a disk was smaller by a factor of several for the "deterministic" approach than that for the "probabilistic" approach.

*Probabilities of collisions of planetesimals with planets b and d in the series MP*

The probabilities of collisions with planet *b* of planetesimals moved from the feeding zone of planet *c* were also calculated for the series *MP*. At integration of the motion for the *MP* series of calculations, planets and planetesimals were considered as point masses. Probabilities of collisions of planetesimals with planets were calculated based on the arrays of orbital elements of migrated planetesimals. Such approach overestimated the probability of a collision of a planetesimal with planets *b* and *d* because in the *MP* calculations planetesimals that should collide with planet *c* (in Fig. 2 for the series *C*, the fraction of such collided planetesimals was typically about 0.3-0.5) still continued to move, and only planetesimals that ejected into hyperbolic orbits or collided with the star were excluded from integration. Actually, a planetesimal that should collide with planet *c* in a few Myr could still move in an elliptical orbit for a greater time interval at the *MP* calculations. The results presented below show that at $e_o$=0.15 the probabilities of collisions of planetesimals with planet *b* are greater by about a factor of two for the *MP* calculations than those for the *C* calculations. The factor is greater for smaller initial eccentricities of planetesimals, as $p_c$ is greater for smaller $e_o$. On the contrary, the probabilities of collisions of planetesimals with planet *c* are smaller by about a factor 1.5-2 for the *MP* calculations than for the *C* calculations (see Appendix B). It also is caused by overestimation of the gravitational influence of planets (e.g. because some retaining planetesimals could pass within the physical radius of a planet). The *MP* calculations allow one to make independent (from the series *C*) estimates of the probabilities of collisions of planetesimals with planets *b* and *d*.

For the series *MP* of calculations made at $1.2 \leq a_{min} \leq 1.7$ AU, the probability $p_b^*$ of a collision of a planetesimal with exoplanet *b* was non-zero in 5 among 18 variants at $e_o$=0.02 and in 3 among 6 variants at $e_o$=0.15. There were 250 initial planetesimals in each variant. Non-zero values of probabilities $p_b^*$ and $p_d^*$ of collisions of a planetesimal with planets *b* and *d* averaged over 250 planetesimals are presented in Table 5. The values of $p_b^*$ and $p_d^*$ averaged over all $N_{mp}$ considered planetesimals and the numbers $N_{b1}$ and $N_{d1}$ of planetesimals with the probabilities of their collisions with planets $b$ and $d$ equal to $p_b^{*1}$=1 are presented in Table 6. In comparison, for the $C$ calculations at 1.2≤$a_{min}$≤1.7 AU and $e_o$=0.15, there were 3 collisions for 3000 initial planetesimals, i.e. $p_b$=10$^{-3}$ was smaller by a factor of 2 than $p_b^*$=2.0×10$^{-3}$ for the $MP$ calculations.

Table 5. Probabilities $p_b^*$ and $p_d^*$ of collisions of a planetesimal with Proxima Centauri $b$ and $d$ for the series $MP$ of calculations at considered time interval $T$=50 Myr. $p_{b0}^*$ is the probability of a collision of a planetesimal with Proxima Centauri $b$ if the eccentricity of the exoplanet's orbit is considered to be equal to 0. $N_b$ and $N_d$ are the numbers of planetesimals (among 250 planetesimals in each variant) which have non-zero probabilities of collisions with Proxima Centauri $b$ and $d$, respectively.

| $a_{min}$ (AU) | $e_o$ | $p_b^*$ | $p_d^*$ | $p_{b0}^*$ | $p_{b0}^*/p_b^*$ | $N_b$ | $N_d$ |
|---|---|---|---|---|---|---|---|
| 1.2 | 0.02 | 0.004 | 0.004 | 0.004 | 1.0 | 1 | 1 |
| 1.3 | 0.02 | 0.00403 | 0.000683 | 0.00402 | 1.0 | 3 | 3 |
| 1.4 | 0.02 | 0.00032 | 0.000143 | 0.000423 | 1.32 | 1 | 1 |
| 1.5 | 0.02 | 1.28·10$^{-5}$ | 0 | 1.78·10$^{-5}$ | 1.39 | 2 | 0 |
| 1.6 | 0.02 | 9.88·10$^{-5}$ | 3.02·10$^{-5}$ | 0.000119 | 1.20 | 1 | 1 |
| 1.4 | 0.15 | 3.64·10$^{-6}$ | 2.58·10$^{-6}$ | 6.22·10$^{-6}$ | 1.71 | 1 | 1 |
| 1.6 | 0.15 | 0.008 | 0.008 | 0.008 | 1.0 | 2 | 2 |
| 1.7 | 0.15 | 0.004 | 0.004 | 0.004 | 1.0 | 1 | 1 |

Table 6. Probabilities $p_b^*$ and $p_d^*$ of collisions of a planetesimal with planets $b$ and $d$ averaged over $N_{mp}$ planetesimals and the numbers $N_{b1}$ and $N_{d1}$ of the planetesimals with the probabilities ($p_b^{*1}$ and $p_d^{*1}$) of their collisions with planets $b$ and $d$ equal to 1 for the $MP$ calculations.

| $e_o$ | $N_{mp}$ | $N_{b1}$ | $N_{d1}$ | $p_b^*$ | $p_d^*$ | $p_b^*/p_d^*$ |
|---|---|---|---|---|---|---|
| 0.02 | 4500 | 2 | 1 | 4.7·10$^{-4}$ | 2.7·10$^{-4}$ | 1.74 |
| 0.15 | 1500 | 3 | 3 | 2.0·10$^{-3}$ | 2.0·10$^{-4}$ | 1 |

For the $MP$ calculations, the values of $p_b^*$=4.7×10$^{-4}$ at $e_o$=0.02 were less than $p_b^*$=2.0·10$^{-3}$ at $e_o$=0.15 by a factor more than 4. This factor (for the difference at $e_o$=0.02 and $e_o$=0.15) can be a little greater for the $C$ calculations because the number of collisions of planetesimals with planet $c$ was greater for smaller $e_o$. Therefore, the ratio of the values of $p_b$ for the $C$ calculations at $e_o$=0.02 to those at $e_o$=0.15 (which were about 10$^{-3}$) can exceed 4 or 5. So the values of $p_b$ for the $C$ calculations at $e_o$=0.02 can be not more than 2×10$^{-4}$, but probably exceed 10$^{-4}$. This estimate does not contradict to the result that there were no collisions of about 3000 planetesimals with planet $b$ for the $C$ calculations at $e_o$=0.02.

The value of $p_b^*$ was calculated for initial eccentricity $e_b$ of the orbit of exoplanet $b$ equaled to 0.11. The arrays of orbital elements of planetesimals obtained at integrations at $e_b$=0.11 were also used for calculations of the probability $p_{b0}^*$ of a collision of a planetesimal with exoplanet $b$ in an orbit with $e_b$=0. In Table 5 for the $MP$ calculations, the ratio $p_{b0}^*/p_b^*$ of probabilities of collisions of planetesimals with Proxima Centauri $b$ for its eccentricity equal to 0 and 0.11 was equal to 1, if $p_b^*$≥0.004. For smaller $p_b^*$, this ratio was in the range between 1.2 and 1.7, i.e., the probability was greater for a smaller eccentricity of planet $b$. In Appendix B it is noted that the corrected probabilities can differ from $p_b^*$ by a factor not more than 1.3. For the $MP$ calculations, collisions of planetesimals with the star were only for two planetesimals (at $e_o$=0.02, $a_{min}$=1.3 AU and $a_{min}$=1.6 AU). The times elapsed before such collisions were equal to 4.0 and 8.9 Myr and were about the same as those for the $C$ calculations.

Exoplanet $d$ was not included in the integrations of motion of planetesimals. However, the probability $p_d^*$ of a collision of a planetesimal with planet $d$ was calculated based on the arrays of orbital elements of migrated planetesimals. For the series *MP* of calculations at 1.2≤$a_{min}$≤1.7 AU, the probability $p_d^*$ was nonzero only for seven (among 24) variants. The value of $p_d^*$ averaged over 1500 planetesimals at $e_o$=0.15 was equal to 2.0×10$^{-3}$. It was greater by a factor of 7.4 than that at $e_o$=0.02. The above values of $p_d^*$ were smaller by a factor of 1.7 than $p_b^*$ at $e_o$=0.02 and were equal to $p_b^*$ at $e_o$=0.15. Therefore, the amount of icy planetesimals delivered to planet *d* could be less than that delivered to planet *b* by a factor less than 2. Almost all contribution to mean values $p_b^*$ and $p_d^*$ for many planetesimals was from those planetesimals for which the calculated probabilities were equal to 1. Those a few planetesimals that have large chances to collide with Proxima Centauri *b* and *d* typically have much greater probabilities of collisions with these exoplanets than with Proxima Centauri *c*. Note that if for some planetesimal $p_b^*+p_d^*$>1, then the probabilities of collisions of these planetesimals with planets *b* and *d* must be divided by $p_b^*+p_d^*$. Such planetesimals have also chances to collide with the star. For a variant of the *MP* calculations with 250 initial planetesimals, not more than three planetesimals (usually one or two) had non-zero probabilities of collisions with exoplanets *b* and *d* (see Table 5).

For the *MP* variants with 250 initial planetesimals, the value of $p_b^*$ exceeded 0.004 at 1.2≤$a_{min}$≤1.3 AU and $e_o$=0.02, and also at 1.6≤$a_{min}$≤1.7 AU and $e_o$=0.15, i.e. for initial orbits of planetesimals not close to the orbit of exoplanet *c*. For the *C* calculations, collisions with planet *b* were at 1.4≤$a_{min}$≤1.5 AU and $e_o$=0.15. The difference in $a_{min}$ could be caused by a small number of considered collisions of planetesimals with planet *b*. Probably, for a large number of planetesimals, collisions of planetesimals with exoplanet *b* can take place at different initial distances from the star. As it is discussed in Appendix B, the probabilities of collisions of planetesimals with planet *c* are smaller and probabilities of collisions with planets *b* and *d* are greater if we consider the mass of planet *c* to be equal to 12$m_E$.

## VARIATIONS OF THE ORBIT OF GROWING PLANET *C* AND THE DELIVERY OF ICY PLANETESIMALS TO PLANETS *b* AND *d*

**The total mass of planetesimals in the feeding zone of planet *c***

As it is seen from Table 3, for the mass of growing planet *c* twice less than the present mass of planet *c* (i.e., at $k_s$=0.5) at *T*=100 Myr, the ratio $p_{cej}=p_c/p_{ej}$ of the number of collided planetesimals to the number of ejected planetesimals was about 1.3-1.5 and 0.5-0.6 at $e_o$=0.02 and $e_o$=0.15, respectively, i.e. $p_c/p_{ej}$ did not differ much from the values (0.8-1.3 and 0.4-0.6) at the mass of planet *c*. For $k_s$=1 and $a_{min}$≤2 AU, the values of $p_{cej}$ were about the same at *T*=10 Myr and *T*=100 Myr (see Fig. 4). At *T*=10 Myr, the ratio of $p_{cej}$ at $k_s$=1 to that at $k_s$=0.5 was about 3 and 2 at $e_o$=0.02 and $e_o$=0.15, respectively. Therefore, the total mass of planetesimals ejected into hyperbolic orbits by planet *c* could be about (3.5-7)$m_E$. Adding the mass of planet *c*, one can conclude that the total mass $m_{fzc}$ of planetesimals in the feeding zone of planet *c* could be about 10$m_E$ and 15$m_E$ at $e_o$=0.02 and $e_o$=0.15, respectively. For the above estimates of $m_{fzc}$, the following formula was used:

$$m_{fzc}=m_c+0.5m_c/p_{cej}, \qquad (1)$$

where $m_c$=7$m_E$ is the mass of planet *c*. Note that the model of growth of planets by accumulation of planetesimals in gas-free environment is considered in this paper. Before this stage, planet *c* could already got some mass, e.g. due to pebble accretion (and this mass should be subtracted from the estimates of the total initial mass of planetesimals). However, noticeable ejection of planetesimals began after growing planet *c* reached $m_E$, and estimates of ejection in formula (1) were mainly for the growth of mass of planet *c* from $m_c$/2. So for my estimates there is no much difference how (via pebble or planetesimal accretion) planet *c* reached the first half of its mass.

**Variation of a semi-major axis of an orbit of a growing planet**

Based on the results of calculations presented above and in Appendix B, below the estimates of the variation in a semi-major axis of an orbit of growing planet $c$ are presented. For the two-body problem for a body-planetesimal (or a planet) with a mass $m$ moving around the star in an orbit with a semi-major axis $a$, the integral of energy is proportional to $m/a$. In this case, summarizing energy of many bodies, it is possible to consider an approximate equation

$$m_{pl0}/a_{pl0}+m_{tot}/\ a_{tb0}=(m_{pl0}+k_{col}m_{tot})/a_{pl} \qquad (2)$$

where $m_{pl0}$ and $a_{pl0}$ are the mass and the semi-major axis of an orbit of a planet at initial moment of time, and $m_{pl}=m_{pl0}+k_{col}m_{tot}$ and $a_{pl}$ are the same values at a later time $t_{end}$, $m_{tot}$ is the total initial mass of bodies with the same value of semi-major axes of their orbits equal to $a_{tb0}$, and $k_{col}$ is the probability of a collision of a body-planetesimal with the planet during $t_{end}$. Formula (2) does not take into account energy loss during collisions, very rare collisions of planetesimals with the star and planet $b$, the distribution of semi-major axes of planetesimals, and nonzero energy of some ejected bodies. In the considered model, the bodies with a total mass equal to $k_{col}m_{tot}$ fall onto the planet, the bodies with a total mass equal to $(1-k_{col})m_{tot}$ were ejected into hyperbolic orbits (their final energy is considered to be 0). It was supposed that all bodies collided with the planet or were ejected into hyperbolic orbits. A semi-major axis of an orbit of a planet could change also due to its interactions with the bodies that have been left in elliptical orbits, but such bodies were not taken into account in the estimates. From (2) denoting $k_{bpl}=m_{tot}/m_{pl0}$, it is possible to obtain

$$a_{pl}/a_{pl0}=(1+k_{col}k_{bpl})/(1+k_{bpl}a_{pl0}/a_{tb0})=[1+p_{cej}(1+k_{bpl})]/[(1+p_{cej})(1+k_{bpl}a_{pl0}/a_{tb0})], \qquad (3)$$

where $k_{col}=p_{cej}/(1+p_{cej})$. The values of the ratio $p_{cej}$ of the number of collided planetesimas to the number of ejected planetesimals are presented in Fig. 4. Different planetesimals have different semi-major axes, but the distribution of initial bodies-planetesimals over semi-major axes of their orbits was not considered in formulas (2)-(4). Denoting $k_m=m_{pl}/m_{pl0}$ and taking into account that the increase in the mass of a planet is $(k_m-1)m_{pl0}=k_{col}k_{bpl}m_{pl0}$, we obtain $k_{bpl}=(k_m-1)/k_{col}$. For the above approximate model at $a_{pl0}=a_{tb0}$, from (3) we have

$$a_{pl}/a_{pl0}=(1+k_{col}k_{bpl})/(1+k_{bpl})=k_mk_{col}/(k_{col}+k_m-1) \qquad (4)$$

At $p_{cej}$ equal to 1 and 0.5, the values of $k_{col}$ are equal to 0.5 and 1/3, respectively. For $k_m=2$ at such values of $k_{col}$, the ratio $a_{pl}/a_{pl0}$ is equal to 2/3 and 1/2, respectively. Therefore, for the considered model at such $p_{cej}$, there was a decrease in a semi-major axis of the orbit of the planet by a factor of 1.5 and 2, respectively. Formulas (3) and (4) for variations of this semi-major axis depend on $p_{cej}$. Based on formula (1), it is possible to obtain the dependence of $p_{cej}$ on the total mass $m_{fzc}$ of planetesimals in the feeding zone of planet $c$. In our case, $m_{fzc}$ is not a free parameter. In the considered model, the planet that increases its mass became closer to the star because of ejection of planetesimals.

At $e_o=0.02$ the semi-major axes $a_{tb0}$ of planetesimals that were not left in elliptical orbits were between 1.2 and 1.8 AU, i.e. the ratio $a_{tb0}/a_{pl0}$ can be considered between 0.8 and 1.2. At $k_m=2$ and $k_{col}=0.5$ using formula (3), one can obtain that the ratio $a_{pl0}/a_{pl}$ is between 1.7 and 1.3 for $a_{tb0}/a_{pl0}$ is between 0.8 and 1.2, and the ratio $a_{pl0}/a_{pl}$ is greater than that obtained at $a_{tb0}/a_{pl0}=1$ by a factor between 1.13 and 0.87. So for the fixed value of $k_{col}$, the consideration of the distribution of $a_{b0}$ does not change much the estimate of $a_{pl0}/a_{pl}$ compared to the estimate at $a_{tb0}/a_{pl0}=1$. As it is seen from Fig. 4, for the planetesimals initially located at the outer range of the feeding zone of a planet, the values of $p_{cej}$ and $k_{col}$ are smaller (and so $a_{pl0}/a_{pl}$ is greater) than for the planetesimals initially located at smaller distances from the star. The values of $p_{cej}$ are less than 0.2 at $e_o=0.02$ and $a_{min}=1.8$ AU, and also at $e_o=0.15$ and $a_{min}\geq 1.8$ AU. On the other hand, at $e_o=0.02$ and $a_{min}=1.2$ AU, $p_{cej}$ was about 2. The fraction of planetesimals collided with a planet on average is smaller for the planet located more far from the star. Therefore, the decrease in the semi-major axis of an orbit of a growing planet caused by "far" planetesimals is greater than that caused by planetesimals located more close to the star. It is possible to suppose that for the growth of the mass of planet $c$ from $3.5m_E$ to $7m_E$ the semi-major axis of its orbit could decrease by at least a factor of 1.5. The greater were eccentricities of planetesimals due to gravitational scattering at their mutual encounters, the greater was such decrease in the semi-major axis of the planet.

At the mass of growing planet $c$ equal to $0.7m_E$ and $T=100$ Myr, the values of $p_{cej}$ were about 10-15 and 2-3 at $e_o=0.02$ and $e_o=0.15$, respectively (see Table 3). Therefore, the variation in the semi-major axis of an orbit of a growing planet at such mass was small, especially for small initial eccentricities of planetesimals' orbits (for small gravitational scattering due to mutual encounters of planetesimals). Due to ejection of planetesimals, a growing planet decreases a semi-major axis of its orbit slower at its smaller mass. The above estimates show only the order of variation of the semi-major axis of an orbit of a planet. For better estimates, studies of more accurate models will be needed. The similar approach can be used for estimates of variarions in orbits of planets in other planetary systems, especially in those with one dominant planet. The above estimates are presented for the model when gas dissapeared. The variations in an orbit of a planet due to its interaction with gas were considered in many papers (see references e.g. in Kanagawa et al, 2018).

**The delivery of planetesimals from the feeding zone of planet $c$ to inner planets**

The total mass of material and water delivered from the feeding zone of Proxima Centauri $c$ to Proxima Centauri $b$ can be considered to be equal to

$$m_{c\text{-}b}=p_b \cdot m_{fzc} \text{ and } m_{ice}=p_b \cdot k_{ice} \cdot m_{fzc}, \qquad (5)$$

respectively, where $m_{fzc}$ is the total mass of the planetesimals beyond the water-ice condensation line that could reach the feeding zone of planet $c$ with time, $k_{ice}$ is the fraction of water in planetesimals, $p_b$ is the probability of a collision of a planetesimal from the feeding zone of planet $c$ with planet $b$.

In the section "Results of calculations", $p_b$ was obtained to be about $2 \cdot 10^{-4}$ at $e_o=0.02$ and about $10^{-3}$ at $e_o=0.15$. It was noted that planetesimals could collide with planet $b$ already at a mass of growing planet $c$ smaller by a factor of 10 than the present mass of planet $c$. At $m_{fzc}=10m_E$ and $p_b=2 \cdot 10^{-4}$ using formula (5), we have $m_{c\text{-}b}=2 \cdot 10^{-3}m_E$. This is a minimum estimate of the total mass of material delivered to planet $b$ from the feeding zone of planet $c$. This estimate corresponds to $e_o=0.02$. For $e_o=0.15$ at $m_{fzc}=15m_E$ and $p_b=10^{-3}$, the estimate of the total mass of material delivered to planet $b$ is $m_{c\text{-}b}=1.5 \cdot 10^{-2}m_E$. Greater values of $e_o$ correspond to the growth of eccentricities of orbits of planetesimals due to gravitational scattering at their mutual encounters. The greater were masses of large planetesimals, the greater could be the growth of eccentricities of orbits of planetesimals. The above estimates have been made for the present mass of planet $b$. The estimates can be smaller if planet $b$ had a smaller mass during its bombardment by planetesimals from the feeding zone of planet $c$. As planet $b$ is closer to the star than planet $c$, we can suppose that it formed faster than planet $c$. As it is seen from Table 6, the probability of collisions of planetesimals with planet $d$ could be less than that with planet $b$ by a factor not more than 1.5. If averaged over all considered planetesimals, which migrated from the feeding zone of Proxima Centauri $c$, the probability of a collision of a planetesimal with Proxima Centauri $b$ or $d$ is greater than the probability of a collision with the Earth of a planetesimal migrated from the zone of the giant planets in the Solar System. The latter probability is typically less than $10^{-5}$ per one planetesimal (Ipatov, 2020). The temperature of planet $c$ is much below zero (it is considered to be about -234 ºC). Therefore, its feeding zone was located farther from the star than the snow line.

Probably, the values of the mass $m_{c\text{-}b}$ of planetesimals delivered from the feeding zone of planet $c$ to planet $b$ for the future computer model, which can take into account the gravitational scattering due to mutual encounters of planetesimals and the growth of planets, will be in the range between the values obtained for our model at $e_o$=0.02 and $e_o$=0.15. For any masses of planetesimals and corresponding growth of eccentricities of their orbits, it is possible to expect that the probability of a collision with planet *b* of the planetesimal formed in the feeding zone of Proxima Centauri *c* can be not less than the probability of the collision of a planetesimal migrated from the feeding zone of the giant planets with the Earth. The ratio of semi-major axes of orbits of planets *c* and *b* equals to 30 and is greater by a factor of 6 than that for Jupiter and Earth. A planet or planets or/and belts of bodies can exist between orbits of planets *c* and *b*. The gravitational influence of such intermediate celestial objects could help some planetesimals moved from the feeding zone of planet *c* to reach planet *b*. However, some migrated planetesimals could collide with such intermediate objects.

**Delivery of water to inner planets**

Not all material collided with a planet could be left in the planet. Some fraction of material of a planetesimal collided with an exoplanet was ejected from the exoplanet. Canup and Pierazzo (2006) found that more than 50 per cent of impactor's water is lost if a planetesimal collides with the Earth with a velocity which is higher than the parabolic velocity by more than a factor of 1.4, and the collision angle is larger than 30°. Bancelin et al. (2017) calculated water loss as a mass fraction of the initial total water after impacts of rocky Ceres-size objects with a planet with the Earth's mass orbiting a solar mass star at 1 AU. In their figure 5, the mass loss equaled to 11, 37, and 60 per cent at collisions with velocities equal to 1, 3, and 5 escape velocities. However, some ejected material could fall back on the exoplanet (Krijt et al., 2017). Orbital velocities of Proxima Centauri *b* and *d* are greater than that of the Earth by a factor of 1.56 and 2.05, respectively. So collision velocities of planetesimals with these planets can be greater than those with the Earth, and more material could be ejected at collisions from planets *b* and *d* than from the Earth. The orbital velocity of planet *c* is about 0.3 of that for the Earth and corresponds to that at 12 AU in the Solar System.

For the Solar System, Raymond et al. (2005) and Haghighipour and Winter (2014) considered that planetary embryos contained 0.1 per cent water at distances from the Sun between 2 and 2.5 AU, and 5 per cent water at distances between 2.5 and 5 AU. The fraction of water in planetesimals could be greater for greater distances from the Sun. Ciesla et al. (2015) analyzed results of several papers devoted to the delivery of volatiles to planets. They concluded that original parent bodies contained more water than seen in the carbonaceous chondrites today. Lodders (2003), Howard et al. (2014), and Ciesla et al. (2015) supposed that solids beyond the water-ice condensation line should be ~50 per cent water in mass. According to Greenberg (1998), the water fraction in a cometary nucleus is about 30 per cent. Rubie et al. (2015) considered that the fraction of water ice in the bodies formed at a distance larger than 6 AU was 20 per cent. Davidsson et al. (2016) concluded that the fraction of ice in comet 67P is within 14 to 33 per cent. Fulle et al. (2017) supposed that, though the volume fraction of water in comet 67P and trans-Neptunian objects is approximately 20 per cent, the bodies born close to the water-ice condensation line contained more water than trans-Neptunian objects. Morbidelli et al. (2012) and Marty et al. (2016) noted that the fraction of water in planetesimals did not exceed 50 per cent. Newcombe et al. (2023) concluded that substantial amounts of water could only have been delivered to the Earth by means of unmelted material, and some material in planetesimals with a diameter greater than 20 km formed during the first 1.5 Myr after the beginning of the Solar System formation (when most radioactive $^{26}$Al had decayed) could be melted. Existence of water in comets, trans-Neptunian objects, and satellites of the giant planets shows that many planetesimals in the feeding zone of the giant planets could by icy. In their studies of maximum water contant from late accretion in TRAPPIST-1 planets, Raymond et al. (2022) assumed that late impactors contained 10 per cent water by mass. Summarizing the above data, we can suppose that the fraction $k_{ice}$ of water in planetesimals in the feeding zone of Proxima Centauri *c* can be between 10 and 50 per cent.

Depending on possible gravitational scattering due to mutual encounters of planetesimals,

the total mass $m_{c-b}$ of material delivered from the feeding zone of planet $c$ to planet $b$ was estimated above to be in the range from $2 \cdot 10^{-3} m_E$ to $1.5 \cdot 10^{-2} m_E$. At the fraction $k_{ice}$ of ice in the planetesimals between 0.1 and 0.5, the values of $m_{ice}=k_{ice} \cdot m_{c-b}$ are between $2 \cdot 10^{-4} m_E$ to $7 \cdot 10^{-3} m_E$. If as Schwarz et al. (2018) we consider $k_{ice}=0.5$, then $m_{ice}$ is between $10^{-3} m_E$ and $7 \cdot 10^{-3} m_E$. Probably, the amount of water in all planetesimals collided with Proxima Centauri $b$ exceeded the mass of water in Earth's oceans ($2 \times 10^{-4} m_E$). The amount of icy planetesimals delivered to planet $d$ was a little less than that delivered to planet $b$. Not all the water delivered to planets $b$ and $d$ was left in these planets.

## CONCLUSIONS

The estimates of the delivery of icy planetesimals from the feeding zone of Proxima Centauri $c$ with a semi-major axis equal to 1.5 AU to inner planets included the studies of the total mass of planetesimals in the feeding zone of planet $c$ and the probabilities of collisions of such planetesimals with planets $b$ and $d$. At integration of the motion of planetesimals from this zone, the gravitational influence of planets $c$ and $b$ and the star was taken into account. At some calculations, the considered time interval reached 1 Gyr. Planetesimals collided with planets or the star or ejected into hyperbolic orbits were excluded from integration. The dependences of the probabilities of collisions of planetesimals with planets and the probabilities of ejection of planetesimals into hyperbolic orbits on initial semi-major axes $a_o$ of orbits of planetesimals were studied for $0.9 \leq a_o \leq 2.3$ AU at their initial eccentricities $e_o$ equal to 0.02 or 0.15. At initial semi-major axes of orbits of planetesimals between 1.2 and 1.7 AU, less than a half of planetesimals still moved in elliptical orbits after 10 Myr, and the ratio of the number of planetesimals collided with exoplanet $c$ to the number of planetesimals ejected into hyperbolic orbits was mainly in the intervals 0.9-1.3 and 0.4-0.6 for $e_o$ equal to 0.02 and 0.15, respectively (both at $T=10$ Myr and at the end of evolution). This ratio is smaller for a greater mass of a planet moving in the orbit of planet $c$ and for greater initial eccentricities of orbits of planetesimals.

Based on estimates of the amount of planetesimals ejected into hyperbolic orbits, it is possible to conclude that during the growth of the mass of planet $c$ from $3.5 m_E$ to $7 m_E$ ($m_E$ is the mass of the Earth) the semi-major axis of its orbit could decrease by at least a factor of 1.5. The total mass of planetesimals in the feeding zone of planet $c$ could be about $10-15 m_E$.

Only one of several hundred planetesimals, which migrated from the feeding zone of Proxima Centauri $c$, reached the orbit of Proxima Centauri $b$ (with a semi-major axis $a_b$=0.0485 AU), and it often also reached the orbit of Proxima Centauri $d$ (with a semi-major axis $a_d$=0.029 AU). The probability of a collision of a planetesimal initially located in the feeding zone of planet $c$ with planet $b$ (averaged over all considered planetesimals) was obtained to be about $2 \cdot 10^{-4}$ at initial eccentricities of orbits of planetesimals $e_o$=0.02 and about $10^{-3}$ at $e_o$=0.15. A lot of icy material and volatiles could be delivered to planets $b$ and $d$. Depending on possible gravitational scattering due to mutual encounters of planetesimals, the total mass $m_{c-b}$ of material delivered from the feeding zone of planet $c$ to planet $b$ was estimated to be in the range from $2 \cdot 10^{-3} m_E$ to $1.5 \cdot 10^{-2} m_E$. At the fraction $k_{ice}$ of ice in the planetesimals between 0.1 and 0.5, the values $m_{ice}=k_{ice} \cdot m_{c-b}$ of the total mass of water in the planetesimals from the feeding zone of planet $c$ collided with planet $b$ are between $2 \cdot 10^{-4} m_E$ to $7 \cdot 10^{-3} m_E$. Probably, the amount of water delivered to Proxima Centauri $b$ exceeded the mass of water in Earth's oceans ($2 \times 10^{-4} m_E$). The amount of material delivered from the feeding zone of planet $c$ to planet $d$ could be a little less (probably by a factor not more than 1.5) than that delivered to planet $b$.

*Acknowledgments* -- The author would like to express his gratitude to Dr. John Chambers and to an anonymous reviewer for very helpful comments that have considerably improved the manuscript. The author acknowledges the support of Ministry of Science and Higher Education of the Russian Federation under the grant 075-15-2020-780 'Theoretical and experimental studies of the formation and evolution of extrasolar planetary systems and characteristics of exoplanets'.

*Data Availability Statement*—The data that support the findings of this study are available from the corresponding author upon reasonable request.

*Editorial Handling*—Dr. Josep M. Trigo-Rodrı́guez

# APPENDIX A. MODELS FOR ESTIMATES OF THE PROBABILITIES OF COLLISIONS OF PLANETESIMALS WITH EXOPLANETS FOR THE *MP* AND *F* SERIES OF CALCULATIONS

For the *MP* and *F* series of integrations, planetesimals and exoplanets were considered as point masses, and the planetesimals that could collide with exoplanets were not excluded from the integration. Therefore, the calculated values $p_{ej}$ and $p_{st}$ of the fractions of planetesimals ejected into hyperbolic orbits and collided with the host star are greater than the corresponding values $p_{ej}^{\#}$ and $p_{st}^{\#}$ for the below models *A* and *B* (see formulas (6) and (7)) which estimate in different ways the decrease in the number of planetesimals due to collisions with exoplanets. This difference is small if the fraction $p_{ex}^{*}=p_{b}^{*}+p_{c}^{*}+p_{d}^{*}$ of planetesimals that could collide with exoplanets (the values calculated based on the arrays of orbital elements of migrated planetesimals are marked by *) is small. For models *A* and *B*, the fraction of planetesimals collided with exoplanets is denoted by $p_{ex}^{\#}=p_{b}^{\#}+p_{c}^{\#}+p_{d}^{\#}$. The ratio of the number of planetesimals moved in elliptical orbits at the considered time to the initial number of planetesimals is denoted by $p_{el}$ for the results of computer simulations of motion of planetesimals. For models *A* and *B*, the similar ratio is denoted by $p_{el}^{\#}$. At the *MP* and *F* series of integrations of motion of planetesimals, collisions of planetesimals with exoplanets were not simulated. Therefore, $p_{st}+p_{ej}+p_{el}=1$. For models *A* and *B*, it was considered that $p_{ex}^{\#}+p_{st}^{\#}+p_{ej}^{\#}+p_{el}^{\#}=1$. Let us denote $k=p_{st}^{\#}+p_{ej}^{\#}+p_{el}^{\#}=1-p_{ex}^{\#}$ and suppose that $p_{ex}^{\#}=k_2 \cdot p_{ex}$, $p_{st}^{\#}=k \cdot p_{st}$, $p_{ej}^{\#}=k \cdot p_{ej}$, and $p_{el}^{\#}=k \cdot p_{el}$. For the **model *A***, supposing that $k_2=k$, it is possible to obtain

$$k=(1+p_{ex}^{*})^{-1}, \; p_{ex}^{\#}=p_{ex}^{*}/(1+p_{ex}^{*}), \text{ and } p_{ej}^{\#}=p_{ej}/(1+p_{ex}^{*}). \tag{6}$$

As the *MP* and *F* series of integrations do not take into account collisions of planetesimals with planets, then $k_2 \leq 1$. For the **model *B***, it was considered that $k_2=1$ and it was obtained that

$$k=1-p_{ex}^{*}, \; p_{ex}^{\#}=p_{ex}^{*}, \text{ and } p_{ej}^{\#}=p_{ej}(1-p_{ex}^{*}). \tag{7}$$

The model *B* shows the upper limit for values of $p_{ex}^{*}$ and the lower limit for $p_{ej}^{*}$.

Analyzing models *A* and *B*, it is possible to conclude that the values of *k* can be between $1-p_{ex}^{*}$ and $(1+p_{ex}^{*})^{-1}$. Examples of the range of the values of *k* for several values $p_{ex}^{*}$ are presented in Table 7. The values of $p_{ej}^{\#}=p_{ej}k$ are estimated to be in the range from $p_{ej}(1-p_{ex}^{*})$ to $p_{ej}/(1+p_{ex}^{*})$, and the values of $p_{ex}^{\#}$ are estimated to be in the range between $p_{ex}^{*}/(1+p_{ex}^{*})$ and $p_{ex}^{*}$. The difference between the values of $p_{ej}^{\#}$ or $p_{ex}^{\#}$ for models *A* and *B* is greater for greater $p_{ex}^{*}$. Based on the values of probabilities $p_{c}^{*}$, $p_{b}^{*}$, and $p_{d}^{*}$, and using similar formulas as those for $p_{ex}^{*}$, we can estimate the values $p_{c}^{\#}$, $p_{b}^{\#}$ and $p_{d}^{\#}$ of the probabilities of collisions of a planetesimal with exoplanets *c*, *b*, and *d*, that take into account the decrease in the number of migrated planetesimals both due to ejections and collisions.

Table 7. The range of values of *k* for several values of $p_{ex}^{*}$.

| Model | $p_{ex}^{*}$ | 0.1 | 0.2 | 0.3 | 0.75 |
|---|---|---|---|---|---|
| A ($k_2=k$) | $k=(1+p_{ex}^{*})^{-1}$ | 0.909 | 0.833 | 0.769 | 0.57 |
| B ($k_2=1$) | $k=1-p_{ex}^{*}$ | 0.9 | 0.8 | 0.7 | 0.25 |

# APPENDIX B. PROBABILITIES OF COLLISIONS OF PLANETESIMALS WITH EXOPLANETS FOR THE *MP* AND *F* CALCULATIONS

**MP calculations**

Initial data for the series *C* and *MP* were the same. For models *A* and *B* (discussed in Appendix A) based on the *MP* calculations, the probability $p_{ex}^{\#}$ of a collision of a planetesimal with exoplanets and the probability $p_{ej}^{\#}$ of an ejection of a planetesimal into a hyperbolic orbit are presented in Table 8. These probabilities were calculated based on the values of $p_{ex}^{*}$ and $p_{ej}^{*}$ averaged over all considered planetesimals with $a_{min}$ from 1.2 to 1.7 AU. The values of the probabilities $p_{ex}^{\#}$ and $p_{ej}^{\#}$ may be between the values obtained for models $A$ and $B$. Based on the data presented in Table 8, one can conclude that the probability $p_{ex}^{\#} \approx p_{c}^{\#}$ of a collision of a planetesimal with exoplanets can be between 0.19 and 0.27 at $e_o$=0.02, and it can be between 0.15 and 0.18 at $e_o$=0.15. The probability $p_{ej}^{\#}$ of an ejection of a planetesimal into a hyperbolic orbit can be between 0.62 and 0.71 at $e_o$=0.02, and it can be between 0.79 and 0.82 at $e_o$=0.15. The values of $p_{ej}^{\#}/p_{ex}^{\#}$ presented in Table 8 show that for the $MP$ calculations the number of planetesimals ejected into hyperbolic orbits is greater by a factor of several (about 4) than the number of planetesimals collided with exoplanets.

Table 8. The probability $p_{ex}^{\#}$ of a collision of a planetesimal with exoplanets and the probability $p_{ej}^{\#}$ of an ejection of a planetesimal into a hyperbolic orbit for models $A$ and $B$ at initial eccentricities of orbits of planetesimals $e_o$=0.02 or $e_o$=0.15. The values of probabilities were calculated based on the values of $p_{ex}^{*}$ and $p_{ej}$ averaged over all considered planetesimals with different values of $a_{min}$ (from 1.2 to 1.7 AU) for the $MP$ series of calculations at $T$=50 Myr.

| Model | $e_o$ | $p_{ex}^{\#}=p_{ex}^{*}/(1+p_{ex}^{*})$ | $p_{ex}^{\#}=p_{ex}^{*}$ | $p_{ej}^{\#}=p_{ej}(1-p_{ex}^{*})$ | $p_{ej}^{\#}=p_{ej}/(1+p_{ex}^{*})$ | $p_{ex}^{\#}/p_{ej}^{\#}$ |
|---|---|---|---|---|---|---|
| A | 0.02 | 0.194 | | | 0.706 | 0.27 |
| B | 0.02 | | 0.273 | 0.624 | | 0.44 |
| A | 0.15 | 0.15 | | | 0.823 | 0.18 |
| B | 0.15 | | 0.18 | 0.795 | | 0.23 |

Table 9. The mean values of $p_c$, $p_{ej}$, and $p_c/p_{ej}$ for the $C$ calculations at $a_{min}$ from 1.2 to 1.7 AU at considered time interval $T$>100 Myr.

| $e_o$ | $p_c$ | $p_{ej}$ | $p_c/p_{ej}$ |
|---|---|---|---|
| 0.02 | 0.43 | 0.458 | 0.94 |
| 0.15 | 0.29 | 0.677 | 0.43 |

For comparison, for the $C$ calculations at $a_{min}$ from 1.2 to 1.7 AU at considered time interval $T$>100 Myr, the mean values of $p_c$, $p_{ej}$, and $p_c/p_{ej}$ are presented in Table 9. For the $MP$ series, the mean values of $p_{ex}^{\#}/p_{ej}^{\#}$ presented in Table 8 were smaller than $p_c/p_{ej}$ for the $C$ calculations by a factor of 3.5 and 2.4 for the model A and by a factor of 2.1 and 1.9 for the model B at $e_o$=0.02 and $e_o$=0.15, respectively. Note that for the $MP$ variants at $e_o$=0.02 the ratio $p_{ex}^{\#}/p_{ej}^{\#}$ at $a_{min}$=1.4 AU was about 2 and was greater by a factor several than at other values of $a_{min}$. Based on Tables 8-9, we can conclude that the ratio $p_c/p_{ex}^{\#}$ was in the range 1.6-2.26 and 1.6-1.9 at $e_o$=0.02 and $e_o$=0.15, respectively. The smaller ratio is for the model B. The $MP$ calculations on average overestimated the ejection of planetesimals into hyperbolic orbits and underestimated their collisions with planet $c$. For the conclusions on accumulation of planets made based on the $MP$ and $F$ calculations it is needed to take into account that these calculations underestimated the probability of collisions of planetesimals with planet $c$ by about a factor of 1.6-2, and overestimated the probability of their collisions with planet $b$ by about a factor of two. According to formulas (6) and (7), we can expect that the probabilities $p_b^{\#}$ and $p_c^{\#}$ could be smaller by a factor not more than $(1+p_{ex}^{*})$ than $p_b^{*}$ and $p_c^{*}$, respectively. In Table 8 $p_{ex}^{*}$<0.3, so $(1+p_{ex}^{*})$<1.3.

**F calculations**

Let us discuss how the values of the mass and the eccentricity of the orbit of planet $c$ can effect on the scattering of planetesimals and the probabilities of their collisions with planets $c$, $b$, and $d$. It can be interesting for studies of formation of different exoplanetary systems, not only Proxima Centauri. For the series $C$, the probabilities of collisions of planetesimals with planets are discussed in the section "Results of calculations" not only for the present mass of planet $c$, but also for the mass of growing planet c equal to 3.5$m_E$ or 0.7$m_E$.

Table 10. Probabilities $p_c^*$, $p_b^*$, and $p_d^*$ of collisions of a planetesimal with exoplanets c, b, and d during time interval T (in Myr) calculated based on the arrays of orbital elements of planetesimals obtained for the series F of calculations (with $m_c$=12$m_E$). $p_{ej}$ and $p_{st}$ are the probabilities of an ejection of a planetesimal into a hyperbolic orbit and of its collision with the host star, respectively. The values of $p_{ej}$ and $p_{st}$ were obtained at integration of motion of planetesimals. $p_{ej}^\#=p_{ej}(1-p_{ex}^*)$ is the estimation of the probability of an ejection of a planetesimal into a hyperbolic orbit for the model B, where $p_{ex}^*=p_b^*+p_c^*+p_d^*$.

| T (Myr) | | | 1 | 1 | 50 | 50 | 50 | 50 | 50 | 50 |
|---|---|---|---|---|---|---|---|---|---|---|
| $a_{min}$ (AU) | $e_c$ | $e_o$ | $p_{ej}$ | $p_c^*$ | $p_{st}$ | $p_{ej}$ | $p_{ej}^\#$ | $p_c^*$ | $p_b^*$ | $p_d^*$ |
| 1.5 | 0 | 0 | 0.36 | 2.4·10⁻⁹ | 0 | 0.612 | 0.612 | 5·10⁻⁸ | 0 | 0 |
| 1.6 | 0 | 0 | 0.61 | 0.096 | 0 | 0.984 | 0.875 | 0.0991 | 0.0094 | 0.0120 |
| 1.5 | 0 | 0.15 | 0.42 | 0.084 | 0 | 0.996 | 0.893 | 0.103 | 0 | 0 |
| 1.6 | 0 | 0.15 | 0.47 | 0.050 | 0 | 1.* | 0.937 | 0.063 | 0 | 0 |
| 1.5 | 0.1 | 0 | 0.63 | 0.015 | 0.016 | 0.976 | 0.939 | 0.024 | 0.0143 | 0.0125 |
| 1.6 | 0.1 | 0 | 0.59 | 3.5·10⁻¹⁰ | 0 | 1.* | 1. | 4.·10⁻¹⁰ | 0 | 0 |
| 1.5 | 0.1 | 0.15 | 0.70 | 0.010 | 0.004 | 0.992 | 0.969 | 0.015 | 0.0083 | 0.0080 |
| 1.5 | 0.1 | 0.15 | 0.66 | 0.020 | 0.02 | 0.968 | 0.924 | 0.027 | 0.018 | 0.011 |
| 1.6 | 0.1 | 0.15 | 0.68 | 0.040 | 0.016 | 0.972 | 0.865 | 0.095 | 0.0146 | 0.0125 |

*Note:* For the series F, the orbital evolution of planetesimals was integrated for the following initial values of orbits and masses of planets: $a_c$=1.489 AU, $e=e_c$, $i_c=e_c/2$ rad., $m_c$=12$m_E$, $a_b$=0.0485 AU, $e_b=i_b$=0, $m_b$=1.27$m_E$. Calculations of $p_d^*$ were made for $a_d$=0.02895 AU, $m_d$=0.29$m_E$, and $e_d=i_d$=0. In variants with $p_{ej}$=1 (marked by *) all planetesimals have been ejected in 7.7 and 5.3 Myr at $e_c$ equal to 0 and 0.1, respectively.

    A few years ago it was considered (Kervella et al., 2020) that the mass of Proxima Centauri c was about 12$m_E$. This mass is greater by a factor of 1.7 than the present estimate of the mass of planet c. Calculations for the F series were similar to those for the MP series, but different mass and eccentricity of the orbit of planet c were considered. Two sets of initial eccentricities and inclinations of the orbit of Proxima Centauri c were considered in the F series: (1) $e_c$=0.1 and $i_c=e_c/2$=0.05 rad and (2) $i_c=e_c$=0. Initial eccentricities $e_o$ of orbits of planetesimals were equal to 0 or 0.15. The semi-major axes of orbits of exoplanets were the same for the series F, MP, and C of calculations. The main difference in initial data for the series F and MP is a greater mass of Proxima Centauri c in the series F. The mass equaled to 12$m_E$ instead of 7$m_E$ in the series MP. For the series F of calculations, the probabilities $p_c^*$, $p_b^*$, and $p_d^*$ of collisions of a planetesimal with exoplanets c, b, and d during time interval T are presented in Table 10. Calculations of $p_d^*$ were made at $e_d=i_d$=0. Table 10 also includes the probabilities $p_{ej}$ and $p_{st}$ of an ejection of a planetesimal into a hyperbolic orbit and its collision with the host star.

    Analysis of Table 10 shows that the values of $p_c^*$ at the series F did not exceed 0.1 and were smaller than those for the series MP. For the series F of calculations at $i_c=e_c$=0 and $e_o$=0.15, the values of the probability $p_c^*$ of a collision of one planetesimal, initially located near the orbit of exoplanet c, with this exoplanet were about 0.06-0.1. For $i_c=e_c/2$=0.05 at $e_o$=0.15, $p_c^*$ was about 0.02-0.04, i.e. it was smaller than for $i_c=e_c$=0. The ratio of the number of ejected planetesimals to the number of planetesimals collided with exoplanets was about 10 or greater. For models A and B (see Appendix A), the minimum estimated values of probabilities $p_c^\#$ and $p_{ej}^\#$ are equal to $p_c^*(1+p_{ex}^*)^{-1}$ and $p_{ej}(1-p_{ex}^*)$, respectively. For variants of Table 10, we have $p_{ex}^*\leq0.084$, exclusive for one variant with $p_{ex}^*$=0.117. At $p_{ex}^*$=0.1, the values of $(1+p_{ex}^*)^{-1}$ and $(1-p_{ex}^*)$ are equal to 0.91 and 0.9, respectively. Therefore, the values of $p_c^\#$ and $p_{ej}^\#$ differ from $p_c^*$ and $p_{ej}$ by not more than 10 per cent. The values of $p_c^*$ for the series F in Table 10 don't exceed 0.1. The fraction of ejected

planetesimals was greater for a greater mass of planet $c$. In comparison, for the *MP* calculations (at a smaller mass of planet $c$), $p_c^*$ was about 0.2-0.3 and 0.15-0.18 at $e_o$=0.02 and at $e_o$=0.15, respectively.

For the series *F* of calculations, the probabilities $p_b^*$ and $p_d^*$ of collisions of a planetesimal with exoplanets $b$ and $d$ were usually greater at $e_c$=0.1 than at $e_c$=0. The ratio $p_{\Sigma bd}=p_{\Sigma b}/p_{\Sigma d}$ of the sums of probabilities of collisions of planetesimals with exoplanets $b$ and $d$ was calculated for all encounters of planetesimals with the exoplanets even if the calculated probability for one planetesimal could considerably exceed 1. Though the values of $p_b^*$ and $p_d^*$ differed from each other by a factor not more than 1.6 (see Table 10), $p_{\Sigma bd}$ varied from 0.06 to 10.5. For the variants *F* of calculations with 250 initial planetesimals in each variant, the numbers $N_b$ and $N_d$ of the planetesimals that had non-zero probabilities of collisions with exoplanets $b$ and $d$ were in the range of 9-20 and 7-18, respectively. These numbers were greater by a factor of several than the values of $N_b$ and $N_d$ (usually equaled to 1 or 2) for the variants *MP* presented in Table 5.

For $e_c$=0.1 and $e_o$=0.15, in all considered variants *F* with 250 planetesimals, $p_b^*$ and $p_d^*$ exceeded 0.008. On average, the values of $p_b^*$ and $p_d^*$ were greater by a factor of several for the series *F* than for the series *MP* of calculations. The obtained results showed that the probability of collisions of considered planetesimals, migrated from the feeding zone of exoplanet $c$, with inner exoplanets was greater for greater values of a mass and an eccentricity of the orbit of exoplanet $c$ and for greater initial eccentricities of orbits of planetesimals.